\documentclass{aa}  
\usepackage{graphicx}
\usepackage{subcaption}
\usepackage{mwe}
\usepackage{booktabs}
\usepackage{tabularx}
\usepackage{natbib}
\usepackage{xcolor}
\bibpunct{(}{)}{;}{a}{}{,}
\usepackage{gensymb}
\usepackage{graphicx}
\usepackage{txfonts}
\usepackage{siunitx}

\usepackage[flushleft]{threeparttable}

\bibpunct{(}{)}{;}{a}{}{,} 

\begin{document}

\title{X-raying the Sco-Cen OB association:\\
The low-mass stellar population revealed by eROSITA}

\author{J. H. M. M. Schmitt\inst{1}, S. Czesla\inst{1}, S. Freund\inst{1}, J. Robrade\inst{1}, P.C. Schneider\inst{1}}

\institute{Hamburger Sternwarte, Universit\"at Hamburg, Gojenbergsweg 112, 21029 Hamburg, Germany\\
           \email{jschmitt@hs.uni-hamburg.de}}

\date{Received \dots; accepted \dots}
\abstract
{We present the results of the first X-ray all-sky survey (eRASS1) performed by the eROSITA instrument
on board the Spectrum-Roentgen-Gamma (SRG) observatory of the Sco-Cen OB association.
Bona fide Sco-Cen member stars are young and are therefore expected to emit X-rays at the saturation level.  The
sensitivity limit of eRASS1 makes these stars detectable down to about a tenth of a solar mass.
By cross-correlating the eRASS1 source catalog with the {\it Gaia} EDR3 catalog, we arrive at a complete 
identification of the stellar (i.e., coronal) source content of eROSITA\ in the Sco-Cen association, and 
in particular obtain for the first time a 3D view of the detected stellar X-ray sources.  Focusing 
on the low-mass population and  placing the optical counterparts identified in this way in a color-magnitude diagram, 
we can isolate the young stars out of the detected X-ray sources and obtain age estimates of the various 
Sco-Cen populations.  A joint analysis of the 2D and 3D space motions, the latter being available only for a
smaller subset of the detected stellar X-ray sources, reveals that the space motions of the 
selected population show a high degree of parallelism, but there is also an additional population of 
young, X-ray emitting and essentially cospatial stars that appears to be more diffuse in velocity space. Its nature is currently unclear.   We argue that with our procedures,  an identification of almost the 
whole stellar content of the Sco-Cen association will become possible once the final {\it Gaia} and eROSITA catalogs 
are available by the end of this decade. We furthermore call into question any source population 
classification scheme that relies on purely kinematic selection criteria.} 

\mail{J.H.M.M. Schmitt, jschmitt@hs.uni-hamburg.de}
\titlerunning{eROSITA Sco-Cen X-ray stars}

\maketitle

\section{Introduction}
\label{sec_intro}

Stars are born not in isolation, but in large groups or associations.  The Sco-Cen OB association is the association nearest to the Sun and can therefore be studied in great detail.  Many and maybe all of the Milky
Way stars were born in associations like this (cf. \citealt{2010bressert}). A detailed understanding of star formation in Sco-Cen
is therefore of substantial relevance to our understanding of star formation in general.  The Sco-Cen association is huge,
extending over the almost entire fourth galactic quadrant from the constellations Crux and Carina over 
Lupus and Centaurus to Scorpius. \cite{2008preib} provided an extensive and excellent overview
covering the vast body of research on the Sco-Cen star-forming region. 

The Sco-Cen OB association consists of approximately 150 B-type stars spread over about 80 degrees in galactic
longitude (cf. Fig.~9 in \citealt{1999zeeuw}), and it is therefore not highly conspicuous on the sky.
\cite{1914kapteyn} realized the parallelism of the tangential motions of these stars, but the reality of 
Sco-Cen OB as a physical entity remained controversial until \cite{1946blaauw} convincingly demonstrated that Sco-Cen OB
is indeed a true moving group. \cite{1946blaauw} also introduced the subdivision of Sco-Cen OB into
three groups, known today as Lower-Centaurus-Crux (LCC), Upper-Centaurus-Lupus (UCL), and
Upper Scorpius (US) in increasing galactic longitude. We also use these groups here, although this
subdivision is to some extent arbitrary.

The definitive study of the Sco-Cen association based on HIPPARCOS data was presented by \cite{1999zeeuw}, who reported (cf. their Tab.~2)
97, 134, and 83 early-type stars in LCC, UCL, and US, respectively.  All of the stars identified
by \cite{1999zeeuw} are of spectral type B (or later), none are of spectral type O.
\cite{2001hoogerwerf} argued that the runaway star $\zeta$~Oph (spectral type O9.5V) 
originated in Sco-Cen and was expelled in the supernova explosion that gave rise to
the pulsar PSR~J1932+1059. If  this interpretation is correct, $\zeta$~Oph would be the last survivor
of the O-star population of Sco-Cen.  The three Sco-Cen subgroups are known to have different ages. US clearly is the youngest subgroup, and the UCL and LCC complexes are older. The controversy
about the ages is significant, because low-mass stars are typically found to have younger ages than higher-mass stars; we refer to
\cite{2020luhman} for references and a recent discussion of this issue.   A popular view interprets the young stellar population
in Sco-Cen as the results of triggered or sequential star formation.  According to \cite{1999preibisch}, star formation began in UCL
approximately 15~Myr ago. After the most massive UCL stars exploded as supernovae, the resulting shock waves
would have triggered star formation in the neighboring regions. For a more extensive discussion of the Sco-Cen complex
and the young local associations, we refer to \cite{2008fernandez}.

To assess the formation history of the Sco-Cen association, a list of members that is as complete as possible clearly is a necessary 
requirement.  However, assigning membership of individual stars to the
Sco-Cen association has always been a problem, and even in the case of the far fewer more massive
stars,  \cite{2011rizzuto} argued that inclusion of more information (in their case, radial velocities) leads to changes in the
detailed membership lists.
The problem of membership assignment becomes even more severe for the far more abundant lower mass stars.
\cite{1999zeeuw} already identified stars of later spectral type, but regardless of the specific membership lists used,
the sensitivity limitations of HIPPARCOS make it  clear that only a very small fraction of the low-mass stellar population 
can be identified with these data.  For the specific case of US, \cite{2008preib}
estimated that 75\% of the US member stars have masses below 0.6~M$_{\odot}$, which accounts
for only 39\% of the total stellar mass, however.   Thus, rather substantial numbers of low-mass stars must
exist in the Sco-Cen region, and numerous studies involving rather different strategies
have been carried out to search for members of this low-mass stellar population (cf. \citealt{2008preib} for an overview).  

Reliable membership lists are obviously a non-negotiable prerequisite for all empirical determinations of the initial mass function,
which in turn provide the link between observations and theory. For a thorough review of IMF research, we refer to \cite{2013kroupa}. \cite{2002preibisch} presented a determination of the IMF in the US region, distinguishing between
the high-mass and low-mass stellar populations.  While the high-mass members are known (but see  \cite{2011rizzuto}),
the low-mass members are much harder to identify, and the area surveyed by  \cite{2002preibisch} only extended
over 9~deg$^2$, a small fraction of the whole US region (cf. Fig.~\ref{sco_cen2D}).

Given the large extent of the full Sco-Cen region, it is clear that any construction of reliable membership lists a challenging
task. As succinctly described by \cite{2019damiani}, the whole Sco-Cen complex comprises a volume of
about 1.6 $\times$ 10$^6$ pc$^3$, in which the {\it Gaia} DR2 catalog lists several hundred thousand stars, depending on the chosen 
selection criteria. As a consequence, the sought-for young stellar 
population comprises only a small fraction of the overall stellar population, and because
a detailed spectroscopic inspection of all stars is not (yet) feasible, efficient search criteria need to be applied.
\cite{2008preib}  provided a detailed overview of the various methods that were applied to
identify the optically rather faint low-mass stellar population.  Spectroscopic investigations typically
focus upon H$\alpha$ emission as a chromospheric (or accretion) indicator and on
lithium as a youth indicator. Photometric investigations in the infrared attempt to
identify the very low mass stellar and brown dwarf population, and finally, kinematic membership
criteria insist that the members' space motions are almost parallel.   
\cite{2008preib}  summarized the identification efforts of Sco-Cen low-mass members (to a large extent
based on ROSAT data ) and
presented separate lists of US, UCL, and LCC members, a large fraction of which are known
X-ray sources. Yet the overall completeness of the membership lists remained unclear, 
given the large sky area covered by the proposed members, given the lower sensitivity of ROSAT
compared to eROSITA, and given the lack of high-accuracy parallaxes and photometry, which are now provided by
{\it Gaia}.
Further searches for low-mass members in US have been performed, for example, by  \cite{2015rizzuto}, and for the
whole Sco-Cen region, \cite{2016pecaut} presented a sample of 493 stars in the mass range
0.7 M$_{\circ}$~$<$ M$_{*}$~$<$ 1.3 M$_{\circ}$, which they argue are ``likely members'' based on
lithium absorption, dwarf or subgiant surface gravities, kinematic distances consistent
with Sco-Cen membership, and HR position consistent with Sco-Cen membership.

With the advent of {\it Gaia} ({\bf G}lobal {\bf A}strometric {\bf I}nterferometer for {\bf A}strophysics, 
but there is no interferometer on board; \citealt{2016gaia}) and eROSITA (\citealt{2021predehl}),  the observational situation 
with respect to the Sco-Cen association and the problem of membership determination has completely changed.
Both {\it Gaia} and eROSITA are ongoing survey missions. Their ``final''
results are likely available only toward the end of the decade.  The enormous potential
of these new data sets can be demonstrated already with existing data, however, and it is the purpose of this paper to demonstrate that the
combination of {\it Gaia}, yielding decisive information on distance, kinematics, and evolutionary status,  and
eROSITA, yielding decisive information on stellar activity, does reveal the full extent of the
expected low-mass stellar population.

The outline of our paper is as follows: We describe the basis for the {\it Gaia} and eROSITA revolutions in Sec.~2 and elucidate the role of 
stellar youth and activity for the assignment of membership in Sco-Cen  in Sec.~3.  The new data and our
analysis are briefly described in Sec.~4, and our new results are presented in Sec.~5.  We conclude
with a discussion and our conclusions in Sec.~6.

\section{The {\it Gaia} and eROSITA revolutions}
\label{sec_rev}

The {\it Gaia} (launched in December 2013) and Spectrum-Roentgen-Gamma (SRG) missions (launched in July 2019) are likely to
revolutionize our understanding of star formation in the Sco-Cen region and in general.
The {\it Gaia} mission is an ESA large mission devoted to the precise measurement 
of positions, magnitudes, distances, and space velocities of more than a billion stars.
A detailed description of the {\it Gaia} hardware, its scientific goals, and in-orbit performance is given by the
\cite{2016gaia}.  So far, the  {\it Gaia} project has released  {\it Gaia} data in three steps:  {\it Gaia} DR1 in 2016,
 {\it Gaia} DR2 in 2018, and an (early) release  {\it Gaia} EDR3 in 2020; further data releases are envisaged in the 
coming  years.
Specifically, {\it Gaia} EDR3 contains positions, proper motions, and parallaxes for about 1.468 $\times$ 10$^9$
sources. Because  {\it Gaia} is expected to operate at least until 2022 and possibly longer, a ``final"
 {\it Gaia} catalog is not likely to be available any time soon.  For the purposes of
this paper, we therefore work with  {\it Gaia} EDR3 data.

The eROSITA instrument ({\bf e}xtended {\bf RO}entgen {\bf S}urvey with an 
{\bf I}maging {\bf T}elescope {\bf A}rray) is the soft X-ray instrument on board
the Russian-German Spectrum-Roentgen-Gamma (SRG) mission.  After its launch from Baikonur,
SRG was placed into a halo orbit around L2, where it is performing a four-year-long
all-sky survey, which started in December 2019.  The eROSITA all-sky 
survey is carried out in a way similar 
to the ROSAT all-sky survey, that is, the sky is scanned in great
circles perpendicular to the plane of the ecliptic.
The longitude of the scanned great circle, that is, the survey rate,  
moves by $\sim$1$^{\circ}$ per day, which means that after half a year the whole sky is covered. This procedure will be
carried out eight times over the mission lifetime.

This paper deals
with data obtained during the first of  the eight eROSITA all-sky surveys.
SRG is scanning all the time with
a scan rate set to 90$^{\circ}$ per hour, that is, one SRG revolution lasts 4~hours.
Given the eROSITA field of view (FOV) of 1$^{\circ}$,
a source will be observed for up to 40~s depending on its location with respect to the precise scan path of eROSITA.
Furthermore, a source near the plane of the ecliptic enters and leaves
the eROSITA scan path within a day and is thus observed six times. Sources at higher
ecliptic latitudes can be observed up to several weeks, but again with the same cadence of
4~hours.  As a consequence, the sensitivity of the eROSITA all-sky survey strongly depends
on ecliptic latitude.  After the end of the eROSITA all-sky survey, the data of all eight individual
surveys will be merged to yield the final eROSITA survey.

The eROSITA instrument is composed of seven X-ray telescopes,
each equipped with their own CCD camera in the respective telescope's focal plane.  
All seven telescopes and cameras are operated independently, but they all observe in parallel.  
Except for some slight differences in the filters, the seven units are identical and thus provide 
a high degree of redundancy of the total system.
The eROSITA energy range is between 0.2~-~8~keV. 
Its spectral resolution approaches 80~eV at an energy of 1~keV, 
which is significantly higher than that obtained by the ROSAT PSPC and 
opens up entirely new scientific investigations. A detailed description of the
eROSITA hardware, mission, and in-orbit performance is presented by \cite{2021predehl}.

The anticipated minimum (point source) flux limit of the ongoing eROSITA all-sky survey is expected to be
about 1~$\times$~10$^{-14}$ erg/cm$^2$/sec and thus exceeds the ROSAT survey  flux limit by a 
factor of at least 20. With increasing ecliptic latitude, the minimum flux decreases even further.
A simple formula to determine the flux limit of eROSITA is given by the expression
$L_X \approx 1 \times 10^{24} \times D_{pc}^2$ (in cgs units), which implies a (minimum)
limiting X-ray luminosity of about 1.5 $\times 10^{28}$ erg/s at a distance of 120~pc. 

{\it Gaia} data have already been used for studies of the Sco-Cen association: 
 \cite{2019damiani} presented a first study of the stellar content of the 
whole Sco-Cen complex based on Gaia DR2 data.
To identify the young low-mass stellar population in Sco-Cen, \cite{2019damiani} carried out a selection both
in proper motion and color-magnitude diagram (CMD) space.
Specifically, \cite{2019damiani}  selected stars in the Sco-Cen region by first demanding that their distance was
below 200~pc, second, demanding that their proper motions or transverse velocities were located
in some specified polygonal region in the proper motion or transverse velocity plane, where already known members
are located, and third, that the
stars be located in again a specially selected area above the main sequence, their so-called "PMS locus".
Using these techniques, \cite{2019damiani}  found 2862, 4511, and 2803 low-mass members
(with approximate BP-RP colors $>$ 1) in the US, UCL, and LCC regions, respectively. If transverse velocity
filtering is applied, these numbers change to 2587, 4077, and 2154 in US, UCL, and LCC, respectively;
cf. Tab.~1 in \cite{2019damiani}.  

A similar effort involving {\it Gaia} DR2 data was undertaken by \cite{2020luhman}, who concentrated on US and 
also used kinematics and photometry primarily from {\it Gaia}
 to isolate candidate Sco-Cen members. In contrast to \cite{2019damiani},  \cite{2020luhman} introduced a mixture
model to separate Sco-Cen members and field stars.  \cite{2020luhman} then combined these selections with a vast body
of optical photometry and spectroscopy to arrive at a ``final'' list of 2020 stars, 1761 of which were classified as ``true'' US members, while
no detailed membership assignment was performed for the UCL and LCC regions.

Neither \cite{2019damiani} nor  \cite{2020luhman} used magnetic activity diagnostics in their attempts to define suitable membership
criteria.  However, with the advent of the eROSITA all-sky survey (eRASS; cf.  \citealt{2021predehl}) 
a new X-ray data set has become available that covers the whole Sco-Cen region with substantially
greater sensitivity and spectral coverage than the  ROSAT all-sky survey carried out in the early 1990s.  
We demonstrate in Sec.~\ref{sec_magact} that first,  young low-mass star Sco-Cen members
must  be vigorous X-ray emitters and second, that the eROSITA sensitivity allows an essentially complete detection of this stellar population.

\begin{figure}  [bht]
\centering  
\includegraphics[scale=0.55]{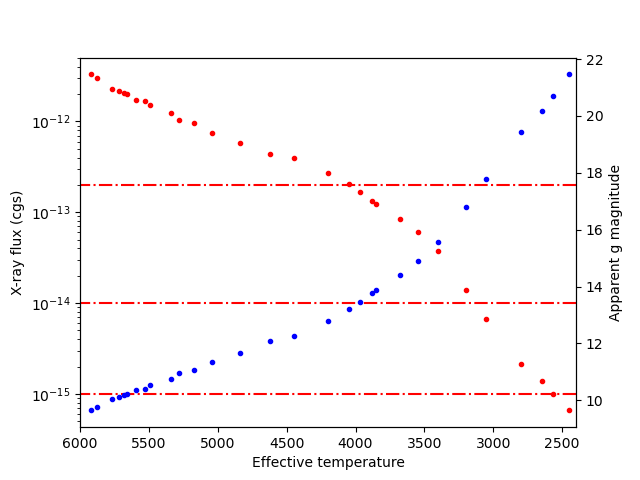}  
\caption{X-ray flux for a saturated coronal source at a distance of 120~pc vs. effective temperature (red data points) and 
apparent {\it Gaia} G magnitude for the same objects (blue data points). The dash-dotted lines refer (from top to bottom) to limiting 
flux levels for the ROSAT all-sky survey, the minimum eROSITA all-sky survey, and to a hypothetical limiting eROSITA 
flux at high ecliptic latitudes. See text for details.}  
\label{gaiaeros}  
\end{figure}

\begin{figure}  [bht]
\centering  
\includegraphics[scale=0.55]{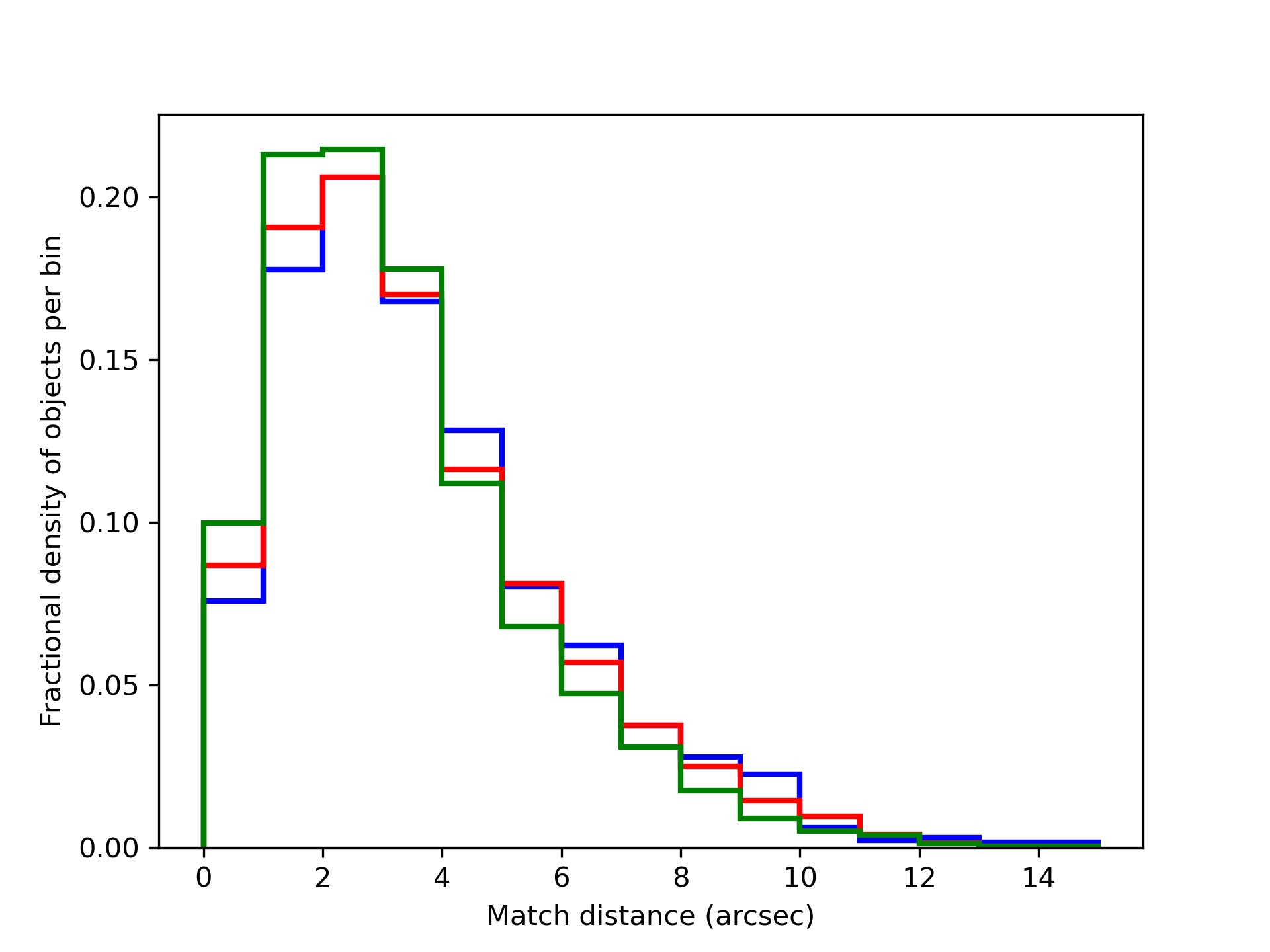}  
\caption{Histogram of position deviations (in arcsec) between eRASS1 and {\it Gaia} EDR3 positions for bona
fide US members (blue histogram), UCL members (red histogram), and LCC members (green histogram).}  
\label{matchdist}  
\end{figure}

\begin{figure}  [bht]
\centering  
\includegraphics[scale=0.55]{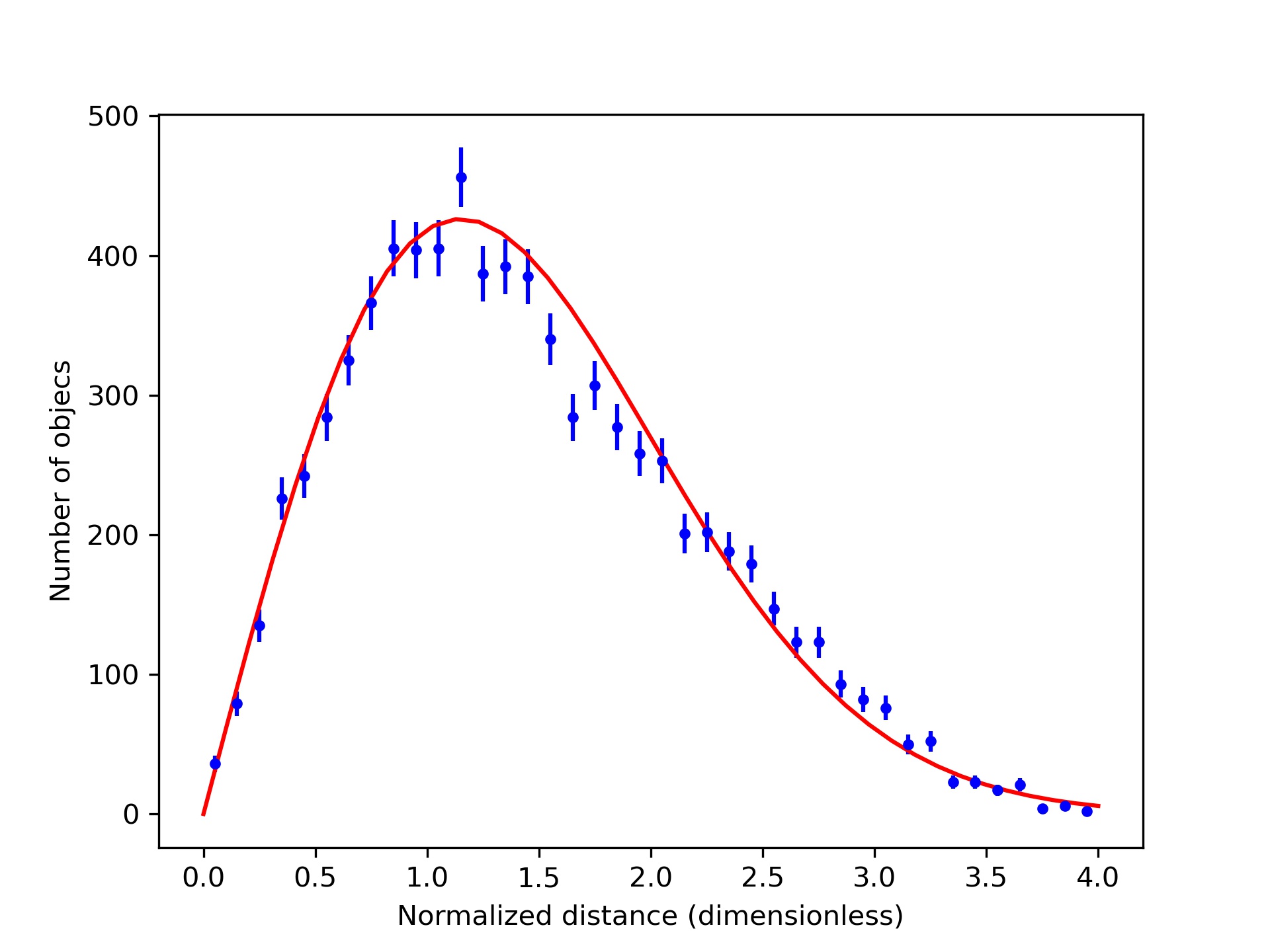}  
\caption{Histogram of normalized position deviations (dimensionless) between eRASS1 and {\it Gaia} EDR3 positions for bona
fide Sco-Cen members (blue data points) and Rayleigh distribution (red curve).}  
\label{normdist}  
\end{figure}

\begin{figure*}  [h]
\centering  
\includegraphics[scale=1.2]{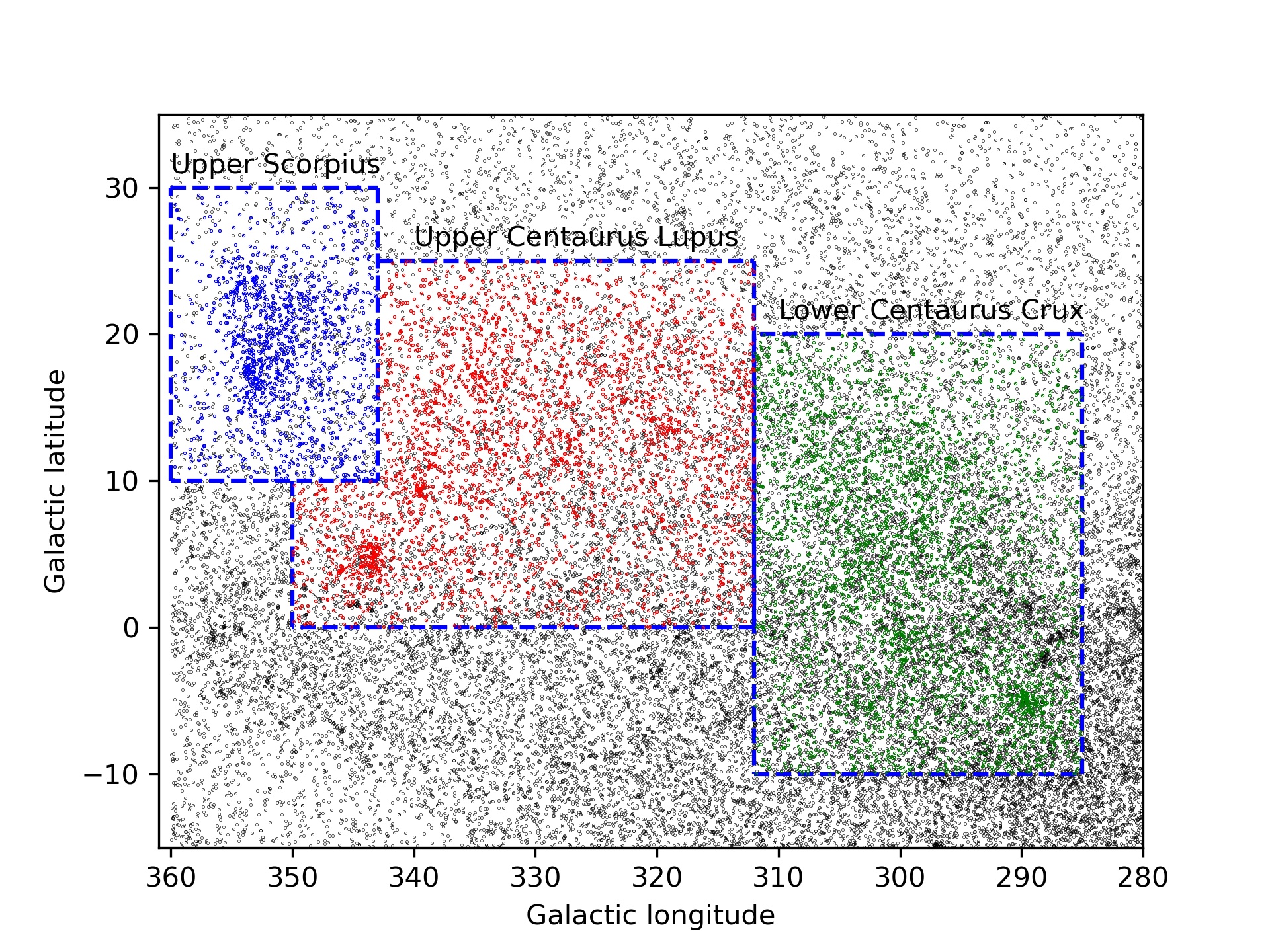}  
\caption{eRASS1 X-ray sources in the fourth galactic quadrant between -10$^{\circ}$<b<30$^{\circ}$. The nominal regions of the Sco-Cen 
subgroups as defined by \cite{2008preib} are indicated.  The colored sources indicates the
sources identified with {\it Gaia} stars in the distance range 60 - 200~pc. Blue sources indicate US member candidates, red sources show UCC member 
candidates, and green sources represent LCC member candidates. Typical eRASS1 position errors are 4 arcsec.}  
\label{sco_cen2D}  
\end{figure*}

\section{Young stars and the role of stellar magnetic activity}
\label{sec_magact}

In the past decades,  overwhelming evidence has been produced that all young stars are indeed very 
rapid rotators (see, e.g., the review by \citealt{2014bouvier}).
According to the rotation-activity paradigm, for late-type stars with outer convection zones, 
rapid rotation is intimately linked with magnetic activity, which manifests itself, among other signatures,
in X-ray emission. The evidence is overwhelming that all stars with outer convection zones are
X-ray sources \citep{1997schmitt}.
The parameter controlling the observed ratio of X-ray ($L_X$) and  bolometric
luminosity ($L_{bol}$) of a 
given star appears to be the  so-called Rossby number,
which is defined as the ratio of rotation period $P_{rot}$ and convective turnover time $\tau_{conv}$.  
Theoretical calculations of $\tau_{conv}$ (see, e.g., \citealt{2010landin}) yield values 
between 10-15 days (for stars with effective temperatures of about 6000~K) to values above 100 days 
and more (for stars with effective  temperatures of 4000~K and below).  

Numerous studies of rotation periods for low-mass stars
have been carried out for open clusters with various ages (cf. the overview by \citealt{2014bouvier} and
references therein).  The initial spin rates (for ages younger than $\sim$ 2~Myr) seem to be distributed
between 1~-10~days, and relatively little rotational spin-down takes place for the first 100~Myr in
the life of a cool star.  Afterward, spin-down is observed, and by the age of the Hyades cluster (625~Myr),
rapid rotation is observed in only a few of its lowest-mass members.  The low-mass member stars of the
Sco-Cen association must have ages not exceeding 20~Myr, for instance, because no O-type stars remain.
Approximately the original
period distribution is therefore expected, implying that the low-mass stars with 
masses below 0.8 M$_{\odot}$ , for instance, invariably have Rossby numbers {\bf far} below unity.

\cite{2011wright} presented a compilation of 
pre-TESS rotational and pre-eROSITA X-ray data for 824 late-type stars.
Fig.~2 in \cite{2011wright} shows that stars with Rossby numbers below $\sim$ 0.1 are 
observed to cluster around a so-called saturation limit of 
approximately $L_X$/$L_{bol}$ $\sim$ 10$^{-3}$, while for stars with larger Rossby numbers,  the observed 
$L_X$/$L_{bol}$-values decrease with increasing Rossby number, leading to the so-called rotation-activity connection. 
While no real physical basis for this saturation limit is known, the saturation limit is well
established empirically, and it remains a challenge to find {\bf coronal} X-ray emitters that do violate the saturation limit.  Coupling the evolution of magnetic activity to a rotational evolution
model, we find that low-mass stars of spectral type K and M stay longer in the saturated regime than higher 
mass stars \citep{2015johnstone}. This emphasizes that X-ray surveys are important to characterize
this population of low-mass stars.  In summary,
the low-mass stellar population in the Sco-Cen association
is expected to be clustered around the X-ray saturation limit. This
saturation limit in our eROSITA data is discussed in Sec.~\ref{sec_xrayprop}.

As a consequence of this saturation limit, the optical counterparts of coronal X-ray sources
in a flux-limited X-ray survey such as eROSITA
will also be (optically) flux limited.  Using specifically the mean stellar color and effective temperature 
sequence compiled by \cite{2013pecaut}, it is straightforward to compute the X-ray luminosity $L_X$ for
a saturated coronal emitter and hence its apparent X-ray flux $f_X$, assuming a distance of
120~pc as appropriate for bulk of the Sco-Cen association as a function 
of effective temperature (red data points in Fig.~\ref{gaiaeros}).
For the same object, we can use its bolometric luminosity, apply the bolometric
correction to the {\it Gaia} band as described by \cite{2018andrae}, and compute its apparent
{\it Gaia} G magnitude, again assuming a distance of 120~pc as a function of effective temperature 
(blue data points in Fig.~\ref{gaiaeros}).   Fig.~\ref{gaiaeros} shows that at this distance of 120~pc, all objects hotter than 3000~K are brighter than G $\sim$ 20 mag and are therefore expected to (eventually)  have measured  {\it Gaia}
parallaxes.  At the same time, these objects would be detectable X-ray sources in X-ray observations
with a limiting flux of  $\approx$ 7 $\times$ 10$^{-15}$ erg/s/cm$^2$, while shallower X-ray surveys
are expected to detect only hotter stars depending on the actual flux limit. 

According to \cite{2012merloni}, eROSITA is expected to reach a limiting flux of at least 1 $\times$ 10$^{-14}$ erg/s/cm$^2$ 
essentially over the whole sky, which would correspond to an effective temperature of $\sim$ 3100~K or a spectral type of  $\sim$ M4V.
In higher ecliptic latitude regions, to which, for example, the LCC region belongs, substantially lower sensitivities can be reached. A limit of  1 $\times$ 10$^{-15}$ erg/s/cm$^2$ , for example, would correspond to an effective temperature of $\sim$ 2500~K or a spectral 
type of  $\sim$ M7V.  The saturation limit is not fulfilled exactly, however. As we demonstrate below (see Fig.~\ref{sco_cenlxlbol} ), there
is substantial scatter, the cause of which is unclear. These
numbers demonstrate, however, that essentially the whole low-mass stellar population in the Sco-Cen region
should be detectable as X-ray emitters by eROSITA.

\begin{figure*}  [h]
\centering  
\includegraphics[scale=0.9]{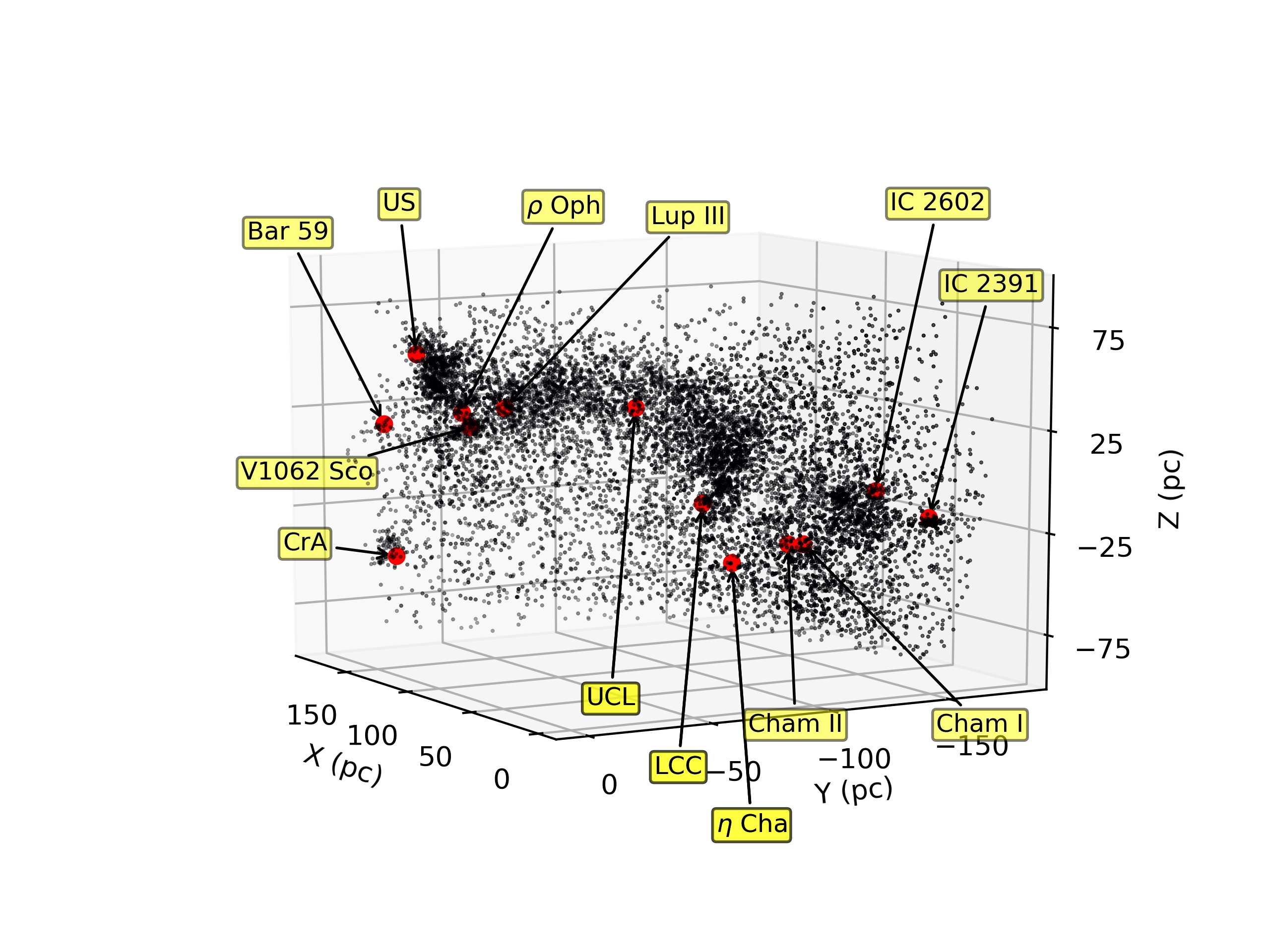}  
\caption{3D view of detected eRASS1 stellar X-ray sources in the fourth galactic quadrant in the (X,Y, Z) coordinate system. For orientation, some individual 
objects or regions in the Sco-Cen association are indicated. See text for details.}  
\label{sco_cen3D}  
\end{figure*}

\begin{figure*}  
\centering  
\includegraphics[scale=1.2]{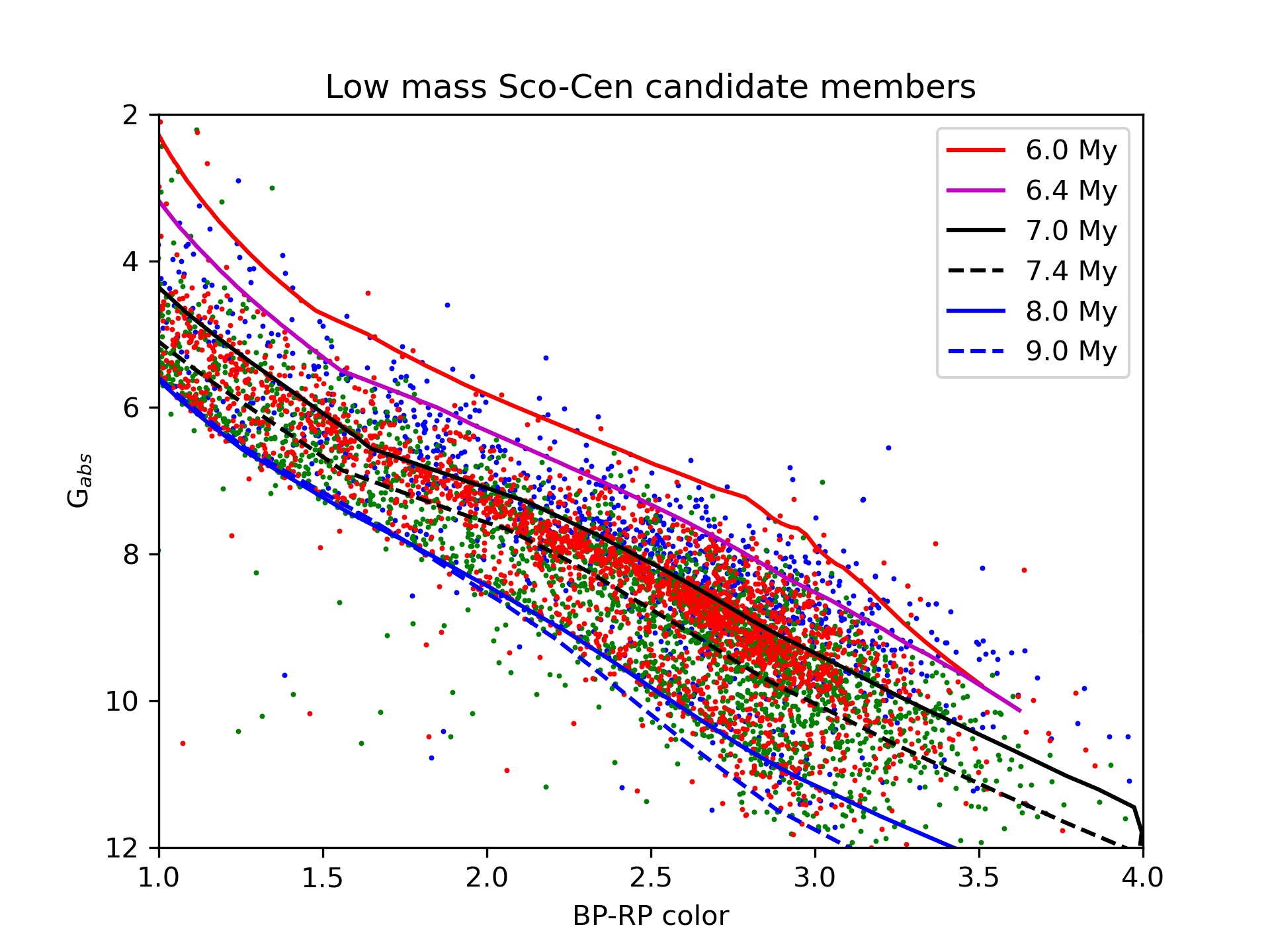}  
\caption{Color-magnitude diagram of low-mass stellar X-ray sources in the US (blue data points), UCL (red data points), and 
LCC (green data points) regions together with PARSEC isochrones (\cite{2012bressan}). The photometry errors are very small.
Parallax errors are typically at the percent level with large variations.}  
\label{sco_cenhrd}  
\end{figure*}

\section{Data and data analysis}
\label{sec_data}

For our eROSITA pilot study of the young stellar population of Sco-Cen, we used
the results of the first of eight planned eROSITA all-sky surveys, using the processing
results of the eSASS pipeline (currently, version 946). A detailed description
of this software and the algorithms we used is given by \cite{2021brunner}.
Briefly,  the eSASS system performs source detection and creates X-ray catalogs for
each sky tile with a size of 3.6$^{\circ}$ x 3.6$^{\circ}$; note that neighboring sky tiles are overlapping.
The X-ray sources in the individual tiles are then merged into one large catalog, in which duplications
in the overlap regions are removed and astrometric corrections are applied.  
The quoted catalog count rates are
fiducial count rates that would be obtained if the source in question had been observed by all seven
eROSITA telescopes on-axis. We note that during the survey, the off-axis angle changes all the time,
and furthermore, not all seven telescopes may deliver useful data at all times due to calibration
activities or malfunction.   

The software performs source detection in the energy bands 0.2~keV~-~0.6~keV,
0.6~keV~-~2.3~keV, and 2.3~keV~-~5.0~keV and also considers a total band 0.2~keV~-~5.0~keV.  Because
stellar coronae tend to be rather soft X-ray emitters compared to active galactic nuclei, we only used the energy 
bands 0.2~keV~-~0.6~keV and
0.6~keV~-~2.3~keV.  We added the count rates obtained in the two lower-energy bands and
used an energy conversion factor of (ECF) of 9 $\times$ 10$^{-13}$ erg cm$^{-2}$ cnt$^{-1}$ to convert
count rates into energy fluxes. This ECF
is appropriate for a thermal plasma emission with a temperature of 1~keV and an absorption column
of 3  $\times$ 10$^{19}$ cm$^{-2}$. No attempt was made to apply individual conversion factors to individual sources
based on their X-ray spectra.  The quoted X-ray fluxes then refer to the band 0.2~keV~-~2.3~keV, which matches the ROSAT band well.

To associate the eROSITA X-ray sources with {\it Gaia} EDR3 entries, we followed the approach described by
\cite{2021schneider}.  The most important ingredient in the matching process is clearly the position, and the
positional offsets between X-ray and optical positions are well known.   The {\it Gaia} EDR3 release contains
1811709771 entries with a very inhomogeneous distribution over the sky.  The {\it Gaia} source
density amounts to 3.4 $\times$ 10$^{-3}$ entries per arcsec$^2$ on average, implying an average number of 0.27 sources
in a randomly chosen circular detection cell of 5" radius. In other words, every fourth match would be a random
match. At low galactic latitudes such as in the Sco-Cen region, the actual source densities are far above average,
which aggravates the confusion problem.

To reduce the number of random matches, it is therefore of paramount importance to reduce the number of 
possible matching candidates.  To this end, we introduced the concept of  ``eligible'' stars, that is, stars that conform to
our preconceived notions of what X-ray emitting stars should look like.   As a first step, we accept as eligible stellar 
counterparts only {\it Gaia} EDR3 entries with a measured parallax (with a signal-to-noise ratio of the parallax S/N $>$ 3), brighter than 
a G magnitude of 18~mag, and  with measured  {\it Gaia} BP-RP colors. Any object violating these criteria may well be an 
X-ray emitter, but it is extremely unlikely
that it is a coronal X-ray emitter.   Confusion with fainter background sources is always a problem, in
particular, when rare source classes are searched for (cf. the example of HR~4289 described by \citealt{1996huensch}). The experience we have gained so far suggests, however,  that this occurs only in very few cases.

It is very important to realize that the optical properties
of a typical stellar eROSITA counterpart are quite different from those of a randomly chosen (even stellar) {\it Gaia} EDR3 entry: In the eFEDS field, for example, we find for the median distance of stellar {\it Gaia} entries a value of  1.1~kpc, that is, the vast
majority of these stars are giants of spectral type G or K  that can optically be detected out to large distances.
However, from the X-ray point of view, these stars are normally quite faint.  For example, the bright
nearby giant Arcturus (V~=~-0.05) is barely detected at X-ray wavelengths \citep{2018ayres} at a level about one hundred times weaker than the 
already weak Sun, while an X-ray source detected at a median distance and at the sensitivity limit would result 
in an X-ray luminosity of $>$ 10$^{30}$ erg/s.  These X-ray luminosities can be attained by rapidly rotating FK Com or RSCVn systems,
but in general, distant giants are  implausible X-ray sources.   To counter the erroneous association of an eROSITA X-ray source 
with {\it Gaia}  eligible stellar background sources,  we therefore devised a selection procedure favoring nearby, brighter, and
red objects over more distant, fainter, and yellower stars. A much more detailed discussion of the whole stellar identification 
problem is provided by \cite{2021schneider}.

To illustrate the positional coincidence of eRASS1 and {\it Gaia} EDR3 positions, we constructed the deviation histograms for the {\it Gaia} and eROSITA
catalog positions shown in Fig.~\ref{matchdist} separately for the
US, UCL, and LCC populations discussed in Sec.~\ref{sec2d3d}.  The positional agreement is clearly excellent; the bulk of the sources is located well 
inside 5 arcsec.   A few sources
have matching distances above 10 arcsec, but all of these sources have large position errors in the eRASS1 data.   
When the positional deviation is normalized by the derived position errors, the distribution of the normalized errors as shown in Fig.~\ref{normdist} (blue data points)
is very well described by a Rayleigh distribution  (red curve in Fig.~\ref{normdist}).    Fig.~\ref{matchdist} and
Fig.~\ref{normdist} therefore clearly show first,  that the bulk of the stellar sources in Sco-Cen is identified correctly, second, that the contamination 
level is very low, and third, that the (positional) completeness of the sample is very high.   Very importantly, we can conclude
that through the excellent positional accuracy of the eROSITA survey data, reliable stellar identifications with low contamination can
also be obtained at low galactic latitudes, where the {\it Gaia} stellar densities are indeed very high.

\section{Results}

\subsection{2D and 3D view of stellar X-ray sources in the Sco-Cen association}
\label{sec2d3d}
In Fig.~\ref{sco_cen2D} we plot all detected eRASS1 point-like
X-ray sources (black data points) in the US, UCL, and LCC regions following the 
convention by  \cite{2008preib} for the definitions
of the respective US, UCL, and LCC regions.  To associate the 
detected X-ray sources with {\it Gaia} EDR3 entries, we followed the procedures described in Sec.~\ref{sec_data}, and following
\cite{2019damiani}, we used a distance screening that accepted only stars located between 60~pc and 200~pc (as calculated from
their  {\it Gaia} measured parallaxes ignoring any errors) as bona fide Sco-Cen member candidates. 

We further note that exposure effects were not removed from Fig.~\ref{sco_cen2D}.
The southern ecliptic pole is located near  $l \sim$ 276$^{\circ}$ and $b \sim$ -31$^{\circ}$, and the increased apparent 
source density in the lower right corner of Fig.~\ref{sco_cen2D}
is caused by the increasing eROSITA exposure in this region.  Proceeding in this fashion, we
identified a total of 7992 matches with  {\it Gaia} stellar counter parts and X-ray sources in the Sco-Cen region 
(1391 in US, 3112 in UCL, and 3419 in LCC), which would constitute a preliminary
sample of X-ray selected bona fide Sco-Cen candidate members. We emphasize that the vast majority of these
candidate members are of low mass, with  {\it Gaia} BP-RP colors in excess of unity. According to
the color tables presented by \cite{2013pecaut}, the {\it Gaia} color BP-RP = 1 corresponds
to a spectral type K0. More specifically, when considering counterparts with  BP-RP $>$ 1,
we find 1108 low-mass US candidate members, 2417  low-mass
UCL candidate members, and 2665  low-mass LCC candidate members. This shows that the percentage of K- and M-type objects 
among the candidate stars is almost 80\%, as would be expected from the initial mass function.

  {\it Gaia} provides precise distances to all the stellar X-ray sources shown in Fig.~\ref{sco_cen2D}.
For visual orientation, we present in Fig.~\ref{sco_cen3D}  a 3D view of all stellar sources
in the fourth$^{}$ galactic quadrant using the usual (X,Y,Z) coordinate system. The positive X-axis points
toward the galactic center, the positive Y-axis in the direction of galactic rotation, and the Sun is located
at (0,0,17). For the height of the Sun above the galactic plane, we refer to the discussion by
\cite{2017karim} for the results of the research of one hundred years for this value. Fig.~\ref{sco_cen3D} shows that the eROSITA-detected X-ray sources are distributed along
a torus-like structure with substantial substructure.  For orientation, some of the better-known
regions are identified in  Fig.~\ref{sco_cen3D}: The positions of the US, UCL and LCC regions
are marked (they are quite large, cf. Fig.~\ref{sco_cen2D} ). Furthermore,
the young clusters IC~2602 and IC~2391, the $\eta$ Cha cluster, and the  V1062~Sco cluster as
well as the star-forming regions in the Chameleon and Lupus constellations, the location of the dark  clouds
associated with Barnard 59, and the  Corona Australis are also indicated.

\subsection{Color-magnitude diagram of stellar X-ray sources in the Sco-Cen association}

Which type of objects are these X-ray selected bona fide Sco-Cen candidate members~?. 
Because we also have the  {\it Gaia} G~magnitude, parallax, and BP-RP color for every X-ray source,
we can construct a color-magnitude diagram (CMD) for these stars, which represents the central result of this study 
and is shown in Fig.~\ref{sco_cenhrd}  for all the US, UCL, and LCC low-mass candidate member stars. We note that
the relevant eROSITA data are provided and described in Appendix \ref{samplestars}.
In addition to the derived CMD positions of the US, UCL, and LCC candidate member stars, we plot
isochrones using the PARSEC suit of isochrones (in version  1.2S, see \citealt{2012bressan}
and  \url{http://www.stev.oapd.inaf.it} ) 
for selected ages in the  range between 10$^{6}$ and 10$^{9}$ yr.   In this context, we note that we did not apply corrections for absorption or binarity to
calculate the CMD positions.
According to \cite{2018andrae}, the reddening vector in a (G--BP-RP) color-magnitude diagram
is characterized by the relation A$_G$ $\sim$ 2 $\times$ E(BP-RP), where E(BP-RP) is 
the BP-RP color excess.
This implies that reddening moves a star very approximately along its isochrone and makes it an even lower-mass star.  \cite{2008preib} provided visual $A_V$ measurements for their US, UCL, and LCC
members that in their great majority are rather low. We therefore expect absorption effects to be 
 low in general, but individual sources near dark clouds or with an almost edge-on disk geometry
 may be substantially affected.
A more severe problem might be unresolved binaries.   Binarity (assuming two equal-mass components)
moves a star by 0.75~mag upward in the CMD, which decreases its apparent age.
Assessing the binarity of all sample stars is currently not possible. 
Finally, there are bound to be systematic errors in the isochrones, especially for the lowest-mass stars
that are difficult to quantify. Our ages are therefore only appropriate in the context of the PARSEC isochrones
we used.
\par
Inspection of Fig.~\ref{sco_cenhrd} shows that the vast majority of the X-ray selected stars
are located between the 10$^{6}$ and 10$^{9}$ yr isochrones.   We point out again
in this context that the only filtering applied was that first, the sources are in the distance range 60 to 200~pc
with a parallax significance of at least 3, and second, they are X-ray detected.  We also note
that a few stars are located up to 3~mag below the 10$^{9}$ yr isochrone and another few points are located above
the 10$^{6}$ yr isochrone.  

The open cluster IC~2602, the southern Pleiades, is located  at the easternmost end of our selected LCC region in 
the constellation Carina, visible as a blob in Fig.~\ref{sco_cen2D} at location ($l \sim$ 289$^{\circ}$, $b \sim$ -5 $^{\circ}$).
The question arises whether it really belongs to LCC.  Various attempts to measure 
the age of IC~2602 have been made; see \cite{2014silaj} for an overview.  For IC~2602, \cite{2014silaj} quoted ages
between 10~Myr and 68~Myr,  while  \cite{1997stauffer} derived an
age of 25~Myr using isochrone fitting to PMS stars.  The most robust estimate is probably the  lithium depletion boundary age of 46~Myr
derived by  \cite{2010dobbie}.    Because the notion of a wave of star formation propagating from
LCC to US is quite popular (cf. \cite{2008preib}, it would be attractive to interpret IC~2602 as one of the older
relics of this process.

Similar considerations apply to the region near  ($l \sim$ 343$^{\circ}$, $b \sim$ 5 $^{\circ}$) in the UCL, which is also
clearly visible as an enhancement in Fig.~\ref{sco_cen2D}.  We associate this group with the young compact moving
group around the star V~1062~Sco that was described by \cite{2018roeser}. V~1062~Sco is a peculiar system of the 
type $\alpha ^{2}$ CVn. The group contains about 60 stars with ages
between 10~Myr and 25~Myr at a distance of 175~pc.  \cite{2018roeser}  also discussed the relation of this moving group
to the UCL complex and instead interpreted it as a subcondensation in velocity space, which is not coeval, however.
 
To include possible members of IC~2602 and V~1062~Sco, we therefore opted for a somewhat generous selection of Sco-Cen member 
candidates in the CMD.
As shown by \cite{2018babusiaux}, the  {\it Gaia} observed main sequence of the Pleiades agrees very well with the
PARSEC 10$^{8}$ yr isochrone, and we therefore selected as low mass Sco-Cen candidate members all eROSITA detected stellar
sources in the respective US, UCL, and LCC regions (see Fig.~\ref{sco_cen2D}), with colors BP-RP $>$ 1.0 (focusing
on K- and M-type stars) that are situated 
between the PARSEC 10$^{6}$ yr and 10$^{8}$ yr isochrones.

\begin{figure}  
\centering  
\includegraphics[scale=0.5]{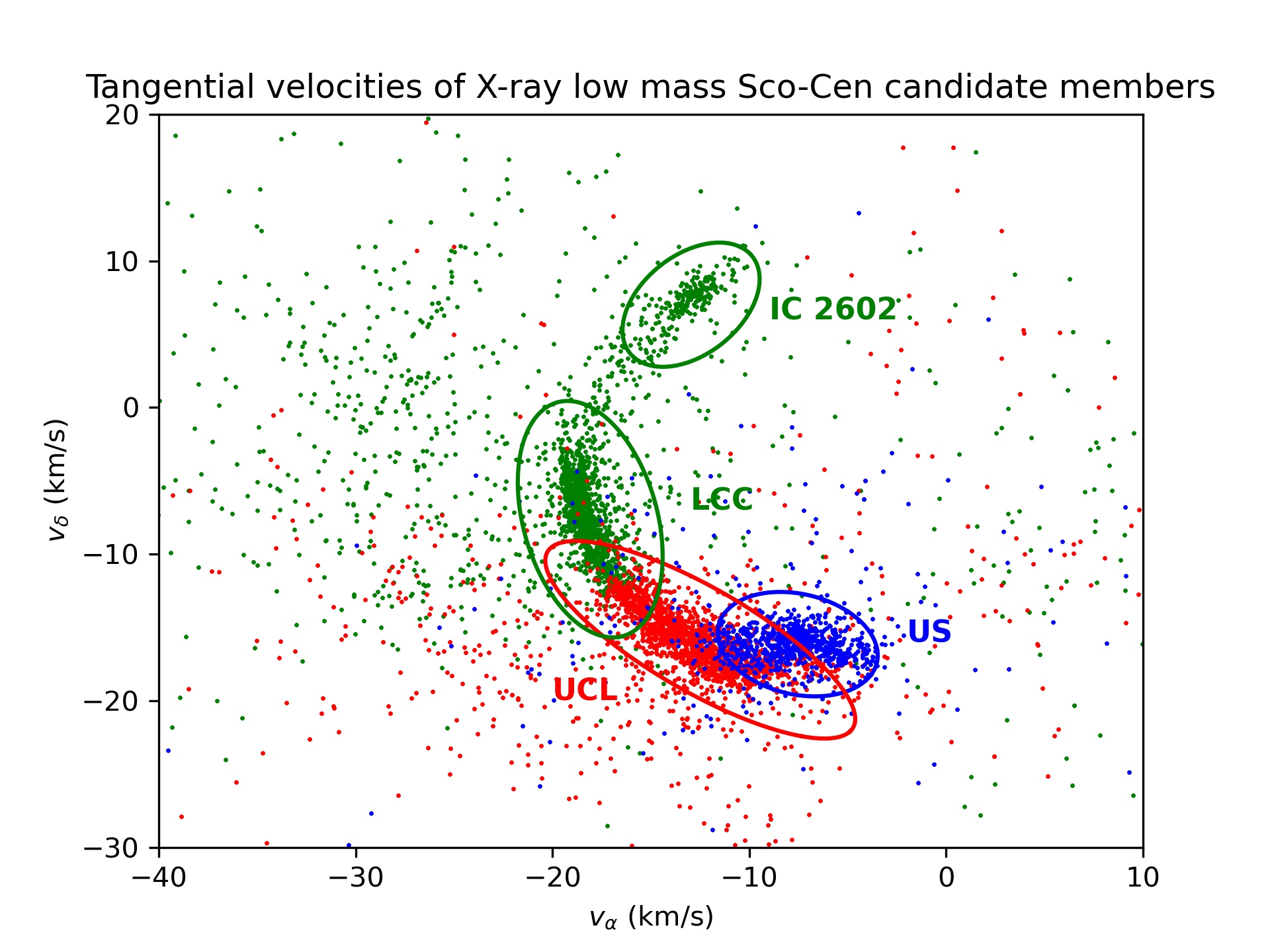}  
\caption{Tangential velocities of low-mass  stellar X-ray sources in the US (blue data points), UCL (red data points), and LCC (green data points) 
regions. The majority of the population is contained in the plotted ellipses, but a population exists that is rather diffuse in
velocity space. See text for details.}  
\label{scocentangvel}  
\end{figure}

\subsection{Kinematics of stellar X-ray sources in the Sco-Cen association}
\label{sec_kin}

Traditionally, kinematics played an important if not decisive role in the membership definition of Sco-Cen, 
while our X-ray based membership assessment so far has ignored kinematics altogether.   Therefore
it behooves us to check the kinematics for our X-ray and CMD selected  Sco-Cen candidate
members.   As is apparent from Fig.~\ref{sco_cen2D}, we are studying rather large regions on the sky extending over almost
90 degrees in longitude, and as a consequence, projection effects can smear out the measured proper motions 
or tangential velocities even for a population that moves exactly in parallel, raising the question as to what 
extent this population actually moves parallel in 3D space. {\it Gaia} currently
does not provide 3D velocity vector information for all Sco-Cen stars, but
we found radial velocities for a the subset of our low-mass Sco-Cen candidates (260 US candidate members, 510 UCL candidate members and 
483 LCC candidate members; see Tab.~\ref{tab1}) in {\it Gaia} EDR3; note that the quoted errors in radial velocity can be quite 
substantial.   It is also clear, however, that for a substantial fraction of our bona fide  members, 
radial velocities and hence 3D vectors are not (yet) available.

\begin{figure}  
\centering  
\includegraphics[scale=0.5]{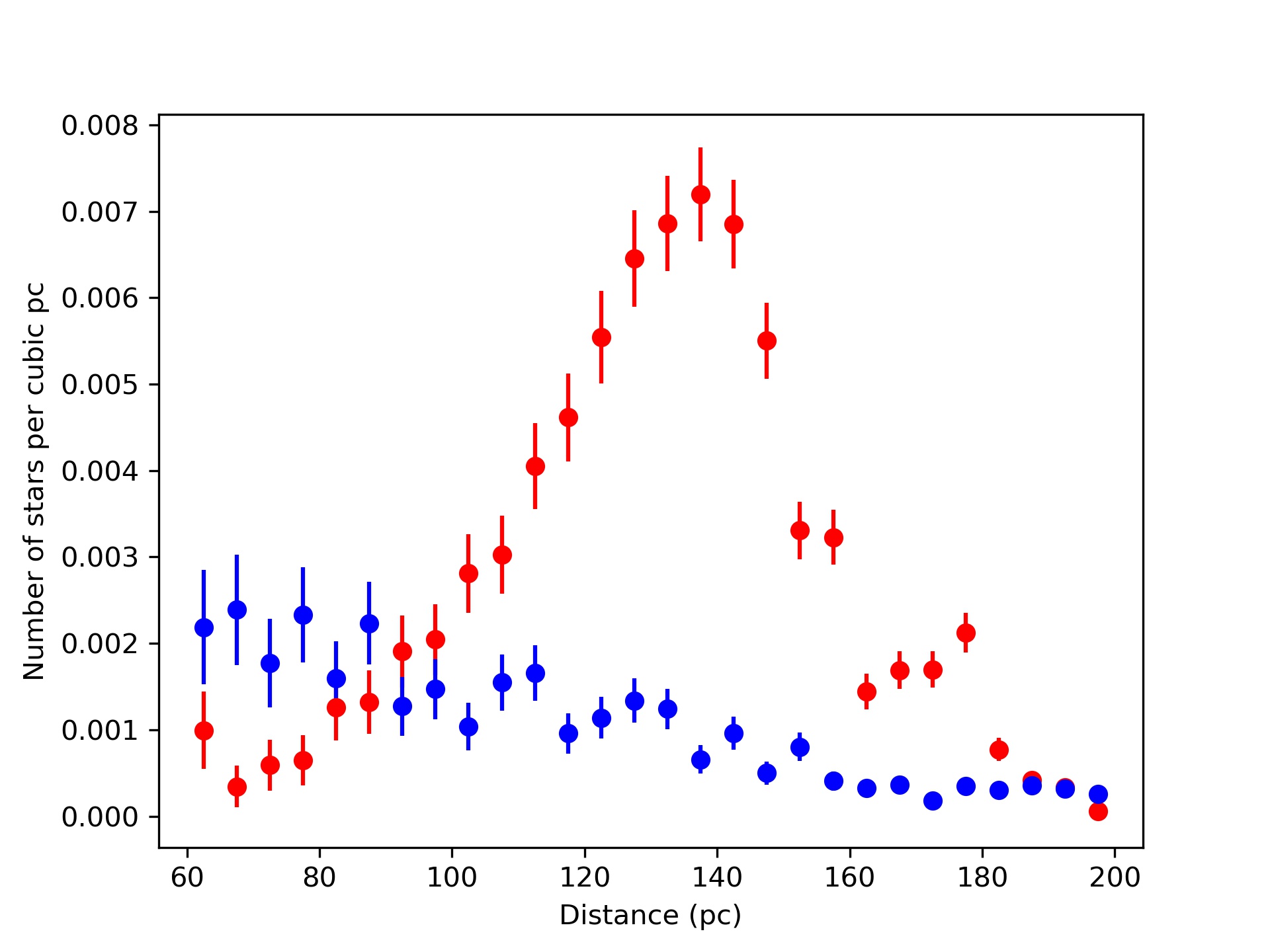}  
\caption{Volume density of young stars (in units of objects per pc$^3$) for the kinematic UCL members (red data points) and cospatial
but kinematic nonmembers (blue data points).}  
\label{voldens}  
\end{figure}

\cite{1987johnson} provided a detailed discussion of velocity vectors and their transformations.  We work with radial
velocities and tangential velocities in the direction of right ascension and declination ($\rho$, v$_{t,\alpha}$,  v$_{t,\delta}$),
where the tangential velocities are computed from the  {\it Gaia} measured proper motions and parallaxes.
As shown by \cite{1987johnson},
the transformation between the observed velocity vector ($\rho$, v$_{t,\alpha}$,  v$_{t,\delta}$) and the true
space motion vector (U,V,W), with component U denoting the velocity component toward the galactic center, component
V the component along the galactic rotation, and component W the component toward north, is given by a linear transformation of the form\\
\begin{equation}
\begin{bmatrix}U\\ V\\ W\\ \end{bmatrix} = \mathbf{T} \cdot \mathbf{A} \cdot \begin{bmatrix}\rho\\ \mathrm{v}_{t,\alpha}\\  \mathrm{v}_{t,\delta}\\ \end{bmatrix} ,
\label{eq1}
\end{equation}
\noindent
where the matrices {\bf T} and {\bf A} only depend on the orientation of the galactic coordinate system with respect to the equatorial
system and the coordinates of the objects.   As discussed above, for the majority of our sources, we are still lacking measurements of $\rho$.
Therefore the individual (U,V,W)  components cannot be computed, but tangential velocities are available for all sample objects.
For a stellar population moving parallel in space, that is, with constant (U,V,W) but dispersed
in distance and position, the distribution of these objects can be found, for example, in the observed (v$_{t,\alpha}$,  v$_{t,\delta}$)-plane,
that is, the tangential velocity plane.  In Fig.~\ref{scocentangvel}  we plot the tangential velocities for all our X-ray and CMD 
selected Sco-Cen candidate members in
our usual color scheme and note a clear clustering in the tangential velocity plane.

Eq.~\ref{eq1} shows that the measured tangential velocities   (v$_{t,\alpha}$,  v$_{t,\delta}$) 
depend linearly on the three space velocity components. Assuming now a source population in parallel motion, but sufficiently distributed in 3D space,
we can (analytically) determine a best-fit 3D velocity vector from the set of measured tangential velocities and the available radial velocities.  Each individual measurement $\rho _i$, v$_{t_{\alpha} ,i}$,  v${_{t,\delta},i}$) depends linearly on (U,V,W) and thus a best fit,
for example, in a least-square sense, can be obtained by minimization.
Furthermore, in practice, the stellar motions under consideration will not exactly be
parallel, but maintain some dispersion from their parent molecular clouds.  To assess the maximum possible extent of a population that moves in parallel in tangential velocity space, we first identified the obvious overdensities in the tangential velocity space
associated with the US, UCL, and LCC population (from 
which we separated
the IC~2602 population) and determined best-fit (U,V,W) velocity vectors for the respective populations by minimizing the distance between
modeled and measured velocity components.  With the best-fit (U,V,W) values obtained in this way (see Tab.~\ref{tab1}), we then carried out a
Monte Carlo calculation, assuming {\it \textup{ad hoc}} a velocity dispersion in each component, and computed the resulting (v$_{t,\alpha}$,  v$_{t,\delta}$) distributions.
To arrive at a simple description of these distributions, we constructed ellipses containing 99\% of the Monte Carlo points. These ellipses are
also shown in the (v$_{t,\alpha}$,  v$_{t,\delta}$) plane  for a population of US, UCL, LCC, and IC~2602 moving in parallel, and all the stars
located inside the ellipses can be considered to move in parallel in space.

Fig.~\ref{scocentangvel} shows that a considerable fraction of data points are located inside the respective ellipses, that is,
the stars form a kinematic entity, but a significant number of data points are located outside, with
the precise numbers  again being provided in Tab.~\ref{tab1}.  If we were to carry out a membership
search based on kinematic criteria alone, the objects located outside would be missed. It is difficult to answer whether membership based on kinematics alone should therefore be rejected, and this requires further study.  
We can definitely say, however, is that the compactness is reduced from US to UCL and LCC, and the data points located outside are also not randomly distributed, but appear to diffuse out 
preferentially along the right ascension direction.   

We further investigated the distance distributions of the populations defined in this way and found them
quite different. In Fig.~\ref{voldens} we show the volume density of young low-mass stars (in terms of objects per pc$^3$) for the
UCL region; the red data points refer to the kinematic UCL members, the blue data points to the kinematic non-members.
Fig.~\ref{voldens}  clearly demonstrates that the population that is compact in velocity space is also spatially far more concentrated
at distances between 120 - 140~pc, while the population that is diffuse in velocity space is spread out almost uniformly over the range of 
distances we considered, and the gradient is probably due to the reduced sensitivity at larger distances.  

\begin{table}[h] 
\centering
\begin{threeparttable}
\begin{tabular}{|l|r|r|r|}  
\hline  
       & US    & UCL  & LCC\\
\hline 
N${_{stellar}}$      & 1335  & 3057 & 3466 \\
\hline 
N${_{late\ type}}$  & 1108 & 2417 & 2665 \\
\hline 
N${_{CMD}}$        &  992  & 2232 & 2371 \\
\hline 
Inside                    &  851  & 1772 & 1697 \\
\hline 
Inside ambiguous   &    51  &     58 &    20\\
\hline 
Outside                 &    90  &   402 &  654\\
\hline 
U (km/s)                &  -5.5   & -7.4   & -9.7\\
\hline 
V (km/s)                & -16.5 & -18.5 &-18.6\\
\hline 
W (km/s)               & -6.8 & -5.6  &-7.0\\
\hline 
N$_{RV all}$        & 260   &  510   &  483 \\
\hline 
 Diffuse                &  9.1\% & 18\% & 28\%\\
 population            &           &          &           \\
\hline  
\end{tabular}  
\caption{\label{tab1} Results of space motion determinations for low-mass Sco-Cen members.}
\vskip 0.25cm
\begin{tablenotes}
\item Notes to Tab.~\ref{tab1}:  N${_{stellar}}$  denotes the
total number of stars, N${_{late\ type}}$ denotes the number of stars with $BP-RP > $1, N${_{CMD}}$  denotes the number
of stars inside the chosen region in the CMD diagram. The row ``inside'' gives the number of stars located inside the respective
ellipse in UV-space (cf. Fig.~\ref{scocentangvel}) centered on the US, UCL, or LCC, respectively, ``inside ambiguous'' gives the number of stars 
inside more than one ellipse, ``outside'' gives the number of stars outside any of the ellipses centered on US, UCL, or LCC.  The
rows U,V,W give the mean space velocity of the ``inside'' stars, N$_{RV all}$  provides the number of stars with Gaia DR2 available
radial velocities, and the row ``Diffuse  population'' provides the estimated percentage of the diffuse population.
The quoted space velocities are with respect to the Sun. No LSR correction has been applied. See text for more details.
\end{tablenotes}
\end{threeparttable}
\end{table}

\begin{figure}  [thb]
\centering  
\includegraphics[scale=0.6]{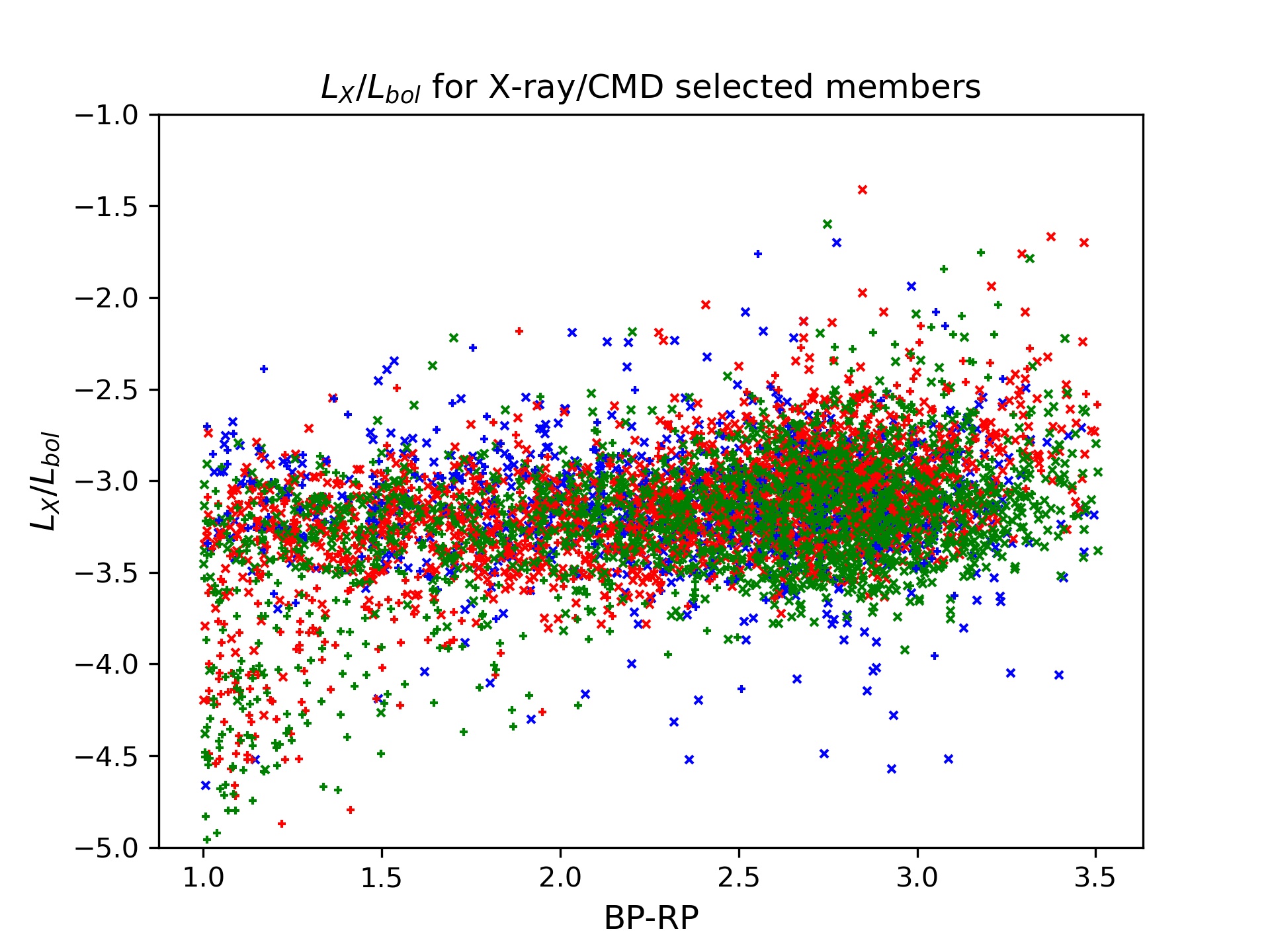}  
\caption{L$_X$/L$_{bol}$ vs. BP-RP color for low-mass kinematic member X-ray sources 
in the US (blue crosses), UCL (red crosses), and 
LCC (green crosses) regions as well as low-mass kinematic nonmembers drawn as 
pluses with the same color code. The saturation limit for stars with BP-RP $>$ 1.5 is well developed in both populations.}  
\label{sco_cenlxlbol}  
\end{figure}

\subsection{X-ray properties of stellar X-ray sources in the Sco-Cen association}
\label{sec_xrayprop}

We now explore the X-ray properties of the various populations identified in Sec.~\ref{sec_kin}.  
Using the eROSITA measured X-ray fluxes and  {\it Gaia} measured distances, we can compute the X-ray luminosities 
for our sample stars.    Furthermore,
using the measured  {\it Gaia} BP-RP colors and the bolometric corrections as described by \cite{2018andrae} and
ignoring any possible absorption effects, we can compute bolometric luminosities L$_{bol}$ for all the sample stars and
proceed to compute their L$_X$/L$_{bol}$ ratios.   For the sake of more clarity, we consider the compact and
diffuse populations as defined in Fig.~\ref{scocentangvel} separately and show in
Fig.~\ref{sco_cenlxlbol}  the measured L$_X$/L$_{bol}$ values as a function of BP-RP color for
low-mass  kinematic members (i.e., the stars located in the respective ellipses shown in Fig.~\ref{scocentangvel})
in the US.  Fig.~\ref{sco_cenlxlbol}  clearly demonstrates the relevance of the saturation limit. The vast majority of the kinematic members and nonmembers are saturated, and only a few of the bluer 
stars in the UCL and LCC regions deviate toward lower values, with the vast majority being kinematic nonmembers.
Some of the (blue) US members show  L$_X$/L$_{bol}$ ratios of 10$^{-4}$ and below, and it is not clear
whether, for example, absorption effects cause this anomaly; the older UCL and LCC populations do not show
such effects.   

Especially the redder stars include a few objects
whose L$_X$/L$_{bol}$ values are higher than  10$^{-3}$.  We point out in this context that all X-ray luminosities
and bolometric luminosities reported in this paper were not measured contemporaneously.  Therefore X-ray flaring
is an obvious possibility to explain anomalously high  L$_X$/L$_{bol}$ values, but the decay times of stellar flares are
typically of the order of hours at best, but (usually) not of days (but see \cite{1996kuerster} for a striking exception).
It remains to be seen whether the further eRASS surveys  will confirm these values. Clearly, with eight surveys at
hand, a characterization of the mean X-ray luminosity of a star is far more robust.
We finally emphasize that in addition to the mere fact of X-ray detection, no
further X-ray properties were used in our identification
process, and therefore Fig.~\ref{sco_cenlxlbol} constitutes a nice example for the validity of our approach because objects
with high L$_X$/L$_{bol}$  values are not discriminated against.

\begin{figure}  [thb]
\centering  
\includegraphics[scale=0.6]{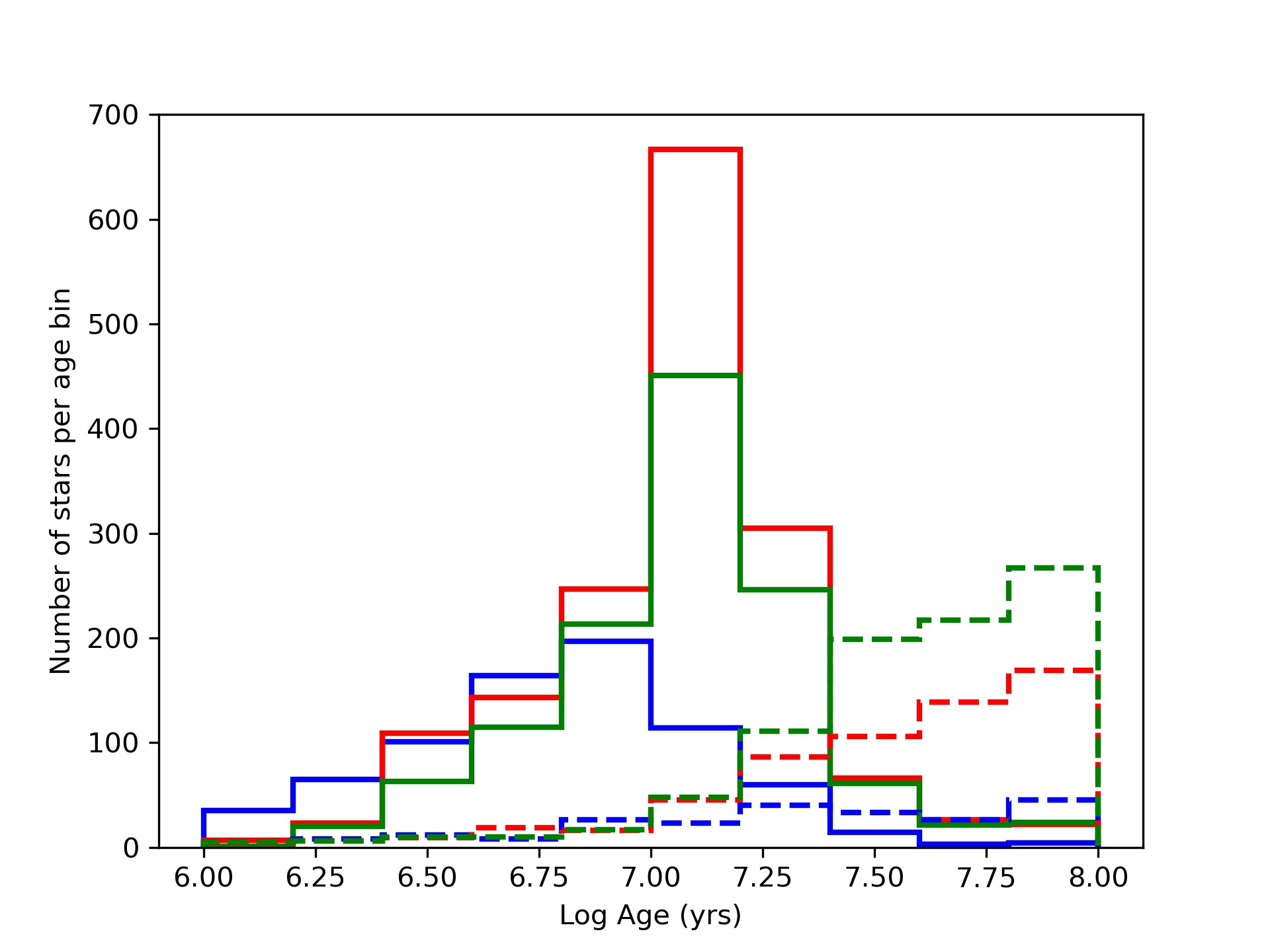}  
\caption{Age distribution histogram for low-mass kinematic member X-ray sources 
in the US (blue histograms), UCL (red histograms), and 
LCC (green histograms) regions. The solid histograms describe the compact population,
and the dashed histograms the diffuse populations. See text for details.}  
\label{sco_cenages}  
\end{figure}

\begin{figure}  [bht]
\centering  
\includegraphics[scale=0.5]{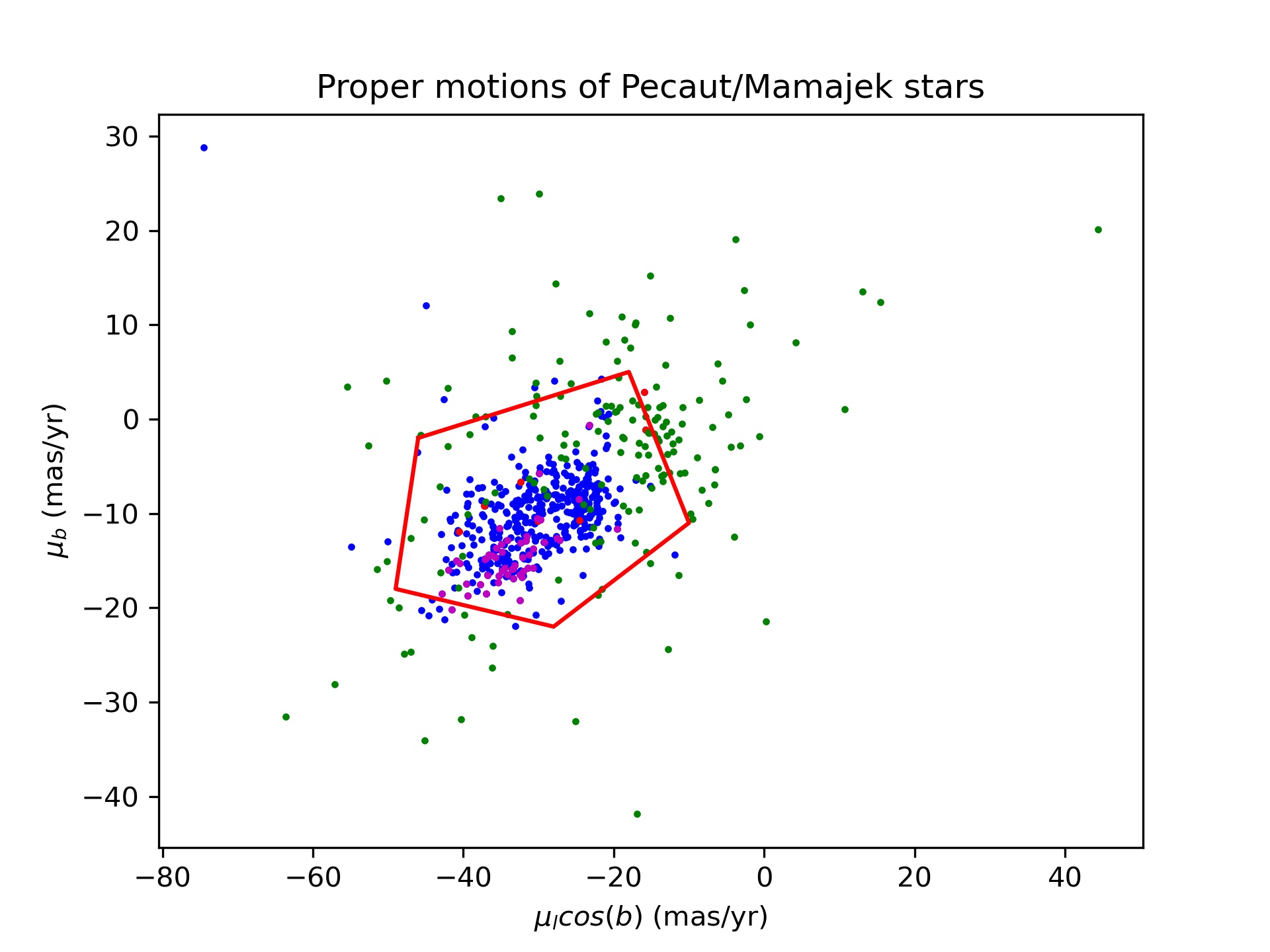}  
\caption{Proper motions (in galactic coordinates) of the likely Sco-Cen association as presented in the study by \cite{2016pecaut}.
Blue data points refer to eRASS1 detected member, green data points to eRASS1 detected nonmembers, red
data points refer to eRASS1 nondetected members, and magenta data points to eRASS1 nondetected nonmembers. The red polygon 
demarcates the selection region for Sco-Cen members chosen by  \cite{2019damiani} to identify Sco-Cen members.} 
\label{scocen_pecaut}  
\end{figure}

\subsection{Ages of stellar X-ray sources in the Sco-Cen association}
\label{sec_age}

Except for kinematics, there are some but not many differences in the X-ray properties of the stars in the compact and
diffuse population.  An obvious question to ask is whether the age distributions of the populations differ and
how this may affect the X-ray emission of the stars.   Because we know the location of each star in the CMD (cf. Fig.~\ref{sco_cenhrd}),
we can determine the age of each star by comparison with the PARSEC isochrones.  
The age distributions determined in this way are plotted in histogram form in Fig.~\ref{sco_cenages} for the compact population in the 
US, UCL, and LCC regions (solid histograms) and  for the diffuse population (dashed histograms),
respectively.   Fig.~\ref{sco_cenages} clearly demonstrates  that the age distributions of these populations
are indeed different.  Only a smaller fraction of the diffuse population has
nominal ages below 10~Myr, the bulk of the derived ages is well above 10~Myr, while in the case
of the compact population, only a tiny fraction has nominal ages above 30~Myr.

\subsection{Comparison to previous work}
\label{sec_comp}

For our selection of Sco-Cen members, X-ray emission does play a decisive role, but we need to determine
how well (or poorly) the X-ray
selection compares to more traditional selection procedures.  To this end, we compared our eROSITA derived results with  previous
studies.  Specifically, we compared our findings to the ROSAT based study by \cite{1998preibisch}, who
obtained lithium spectroscopy for X-ray selected members {\bf and} proper motion selected members without X-ray detections.
Furthermore, we compared our results to the membership lists derived by \cite{2016pecaut}, \cite{2019damiani}, and \cite{2020luhman}.  
As we discussed in Secs.~\ref{sec_intro} 
and \ref{sec_rev}, previous membership lists of Sco-Cen members have been
published that were based on rather different selection methods and on different data sets, which we discuss in turn 
below.  \cite{1998preibisch}
and  \cite{2016pecaut} only used pre-{\it Gaia} data, while \cite{2019damiani} and \cite{2020luhman} used {\it Gaia} DR2 data 
to various extents.

\subsubsection{Comparison to the membership lists derived by \cite{1998preibisch} }

\cite{1998preibisch} selected a 160 square degree region in US (cf. their Fig.~1) and primarily used ROSAT all-sky survey data (RASS, in addition
to some deeper pointed ROSAT data) in that area.  \cite{1998preibisch}  reported  a total of 606 X-ray sources in their specified survey region and obtained spectroscopy for 69 RASS  sources (out of 130 selected candidate members). In addition,  \cite{1998preibisch}  obtained spectroscopy
for 115 stars that they deemed to be Sco-Cen members based on their proper motions but remained undetected at X-ray  wavelengths.
In 39 (out of 69) stars,  \cite{1998preibisch}  detected a lithium equivalent width above the threshold value (cf. their Fig.~3), while 
none of the X-ray quiet proper motion candidate members showed significant lithium absorption.  Comparing the RASS performance to 
eROSITA, a total of 3556 sources are found in the same area (already in eRASS1), out of which 868 are bona fide low-mass US members. 
Interestingly, in eRASS we do not recover all of the 69 RASS sources reported by  \cite{1998preibisch}. For 14 of them, no eRASS1 detection could
be obtained, while only 2 of the 115 proper motion candidate members resulted in an eRASS1 detection.    This clearly shows that already eRASS1 supersedes
the RASS numbers considerably, and the final eRASS8 data will again be much deeper and will in addition allow us to address long-term variability in
the X-ray sources in the Sco-Cen region.

\begin{figure}  [bht]
\centering  
\includegraphics[scale=0.5]{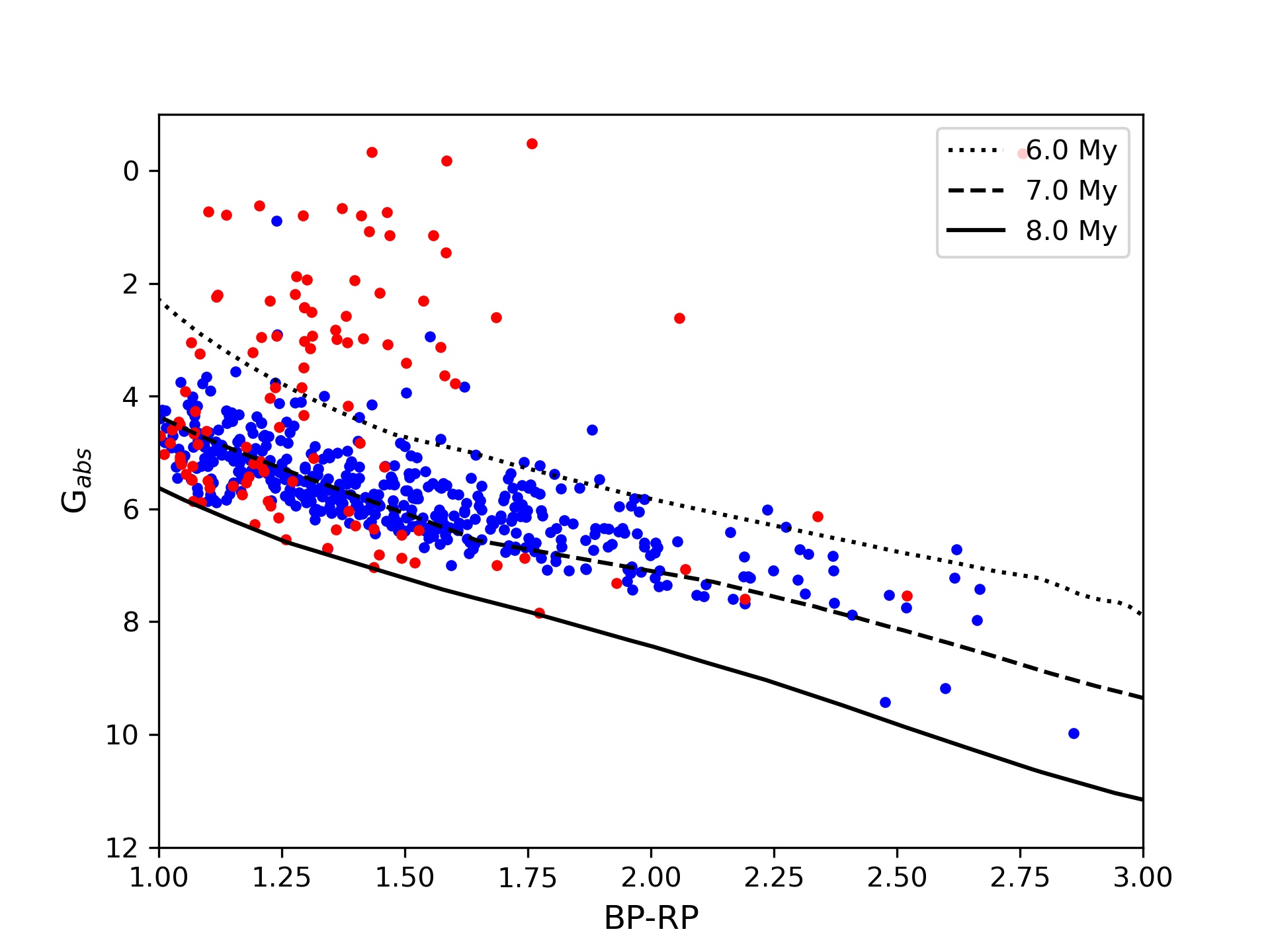}  
\caption{CMD diagram of Sco-Cen members (blue data points), accepted  by \cite{2016pecaut} and detected in eRASS1, as well as
stars that were rejected by \cite{2016pecaut}  but were also detected in eRASS1 (red data points). PARSEC isochrones (\cite{2012bressan})
for the ages 1, 10, and 100~Myr are also shown.} 
\label{scocen_pecaut_cmd}  
\end{figure}
\subsubsection{Comparison to the membership lists derived by \cite{2016pecaut} }
\label{sec_pecaut}

In another pre-{\it Gaia} study, \cite{2016pecaut} presented a sample of 493 low-mass stars in the mass range
0.7 M$_{\circ}$ $<$ M$_{*}$ $<$ 1.3 M$_{\circ}$, which they classified as ``likely members'' of the Sco-Cen association 
based upon the criteria lithium absorption, dwarf or subgiant surface gravities, kinematic distances, 
and HRD positions consistent with Sco-Cen membership (see Tab.~7 in \citealt{2016pecaut}) as well as a list of 180 stars that they rejected as likely members because one or more of the membership criteria were violated 
(see Tab.~8 in \cite{2016pecaut}).    In this context, however, we note that the input sample selection
was based upon the existence of RASS sources at a distance of 40", proper motions in some defined areas, and
magnitude limits 7 $<$ J $<$ 11; see section 2 by \cite{2016pecaut} for a detailed discussion.
For better comparability with our analysis, we cross-matched these sources with {\it Gaia} EDR3,
focused on stars redder than BP-RP = 1, and finally obtained
471 accepted and 169 rejected members from the lists presented by \cite{2016pecaut}.   
A cross-check with the eRASS1 X-ray sources shows
that out of the 471 listed likely members, 464 are detected in the framework of eRASS1, that is, the X-ray
selection yields 98.5 \% of the likely members proposed by \cite{2016pecaut}. However, among the stars rejected 
by \cite{2016pecaut}, we find 115 X-ray detections, that is, an eRASS1 detection rate of 68\%. This is still high, but far lower than 
the detection rate for the likely members.   This discrepancy is somewhat surprising because 
X-ray selection did play some role in the sample definition applied by  \cite{2016pecaut}. The
X-ray properties of the stellar sources were not further used in the ensuing member selection process, however.

To further examine the differences between accepted and rejected members (according to \citealt {2016pecaut}), it is instructive
to examine the proper motion diagram of these objects with {\it Gaia} data. {\it Gaia} EDR3 does not provide distances and
proper motions for all of the objects listed by  \cite{2016pecaut}, which were therefore left out from this comparison.  
In Fig.~\ref{scocen_pecaut} we present this comparison, showing the proper motion diagram for 
the likely and eROSITA detected (blue data points) and rejected, but eROSITA detected stars (green points). We also indicate the
stars that are classified as members but were not detected (red data points) and those that were rejected as members and remain without an eROSITA
detection (magenta data points).   Fig.~\ref{scocen_pecaut} shows that the proper motions of the vast majority of the
\cite{2016pecaut}  are inside the red polygon (see Sec.~\ref{sec_rev}), as expected; this will be further discussed in 
Sec.~\ref{sec_damiani}.  A few of the \cite{2016pecaut}  member stars are located far away from this region, however.
Furthermore, a substantial fraction of the stars rejected by \cite{2016pecaut} as members (green data points in Fig.~\ref{scocen_pecaut} )
because one or more of their selection criteria were violated (see Tab. 8 in \cite{2016pecaut}) are
located inside the red polygon and are detected as X-ray sources.

We finally constructed a CMD for the eROSITA detected  accepted (blue data points) and rejected (red data points) members 
(according to \citealt {2016pecaut}) using 
the {\it Gaia} measured parallaxes, magnitudes, and colors. We show this in Fig.~\ref{scocen_pecaut_cmd} together with the PARSEC isochrones for the ages 1,10, and 100~Myr.  
Fig.~\ref{scocen_pecaut_cmd} demonstrates that the vast majority of  accepted members are indeed located between the
1 and 100~Myr isochrones.   For the  rejected eROSITA detected  members, the situation is less clear: quite 
a few of them are located far above the 1~Myr isochrone and can therefore not be normal stars, while the great majority of the
stars inside the 1 and 100~Myr isochrones have a BP-RP color below 1.5 and probably are of spectral type between K0 and KV.
In this spectral range, the Li equivalent widths measured for cluster stars are largest and the equivalent width
distributions may even partially overlap. This makes the measured  Li equivalent widths a less reliable indicator for Sco-Cen membership.

\begin{figure}  [bht]
\centering  
\includegraphics[scale=0.5]{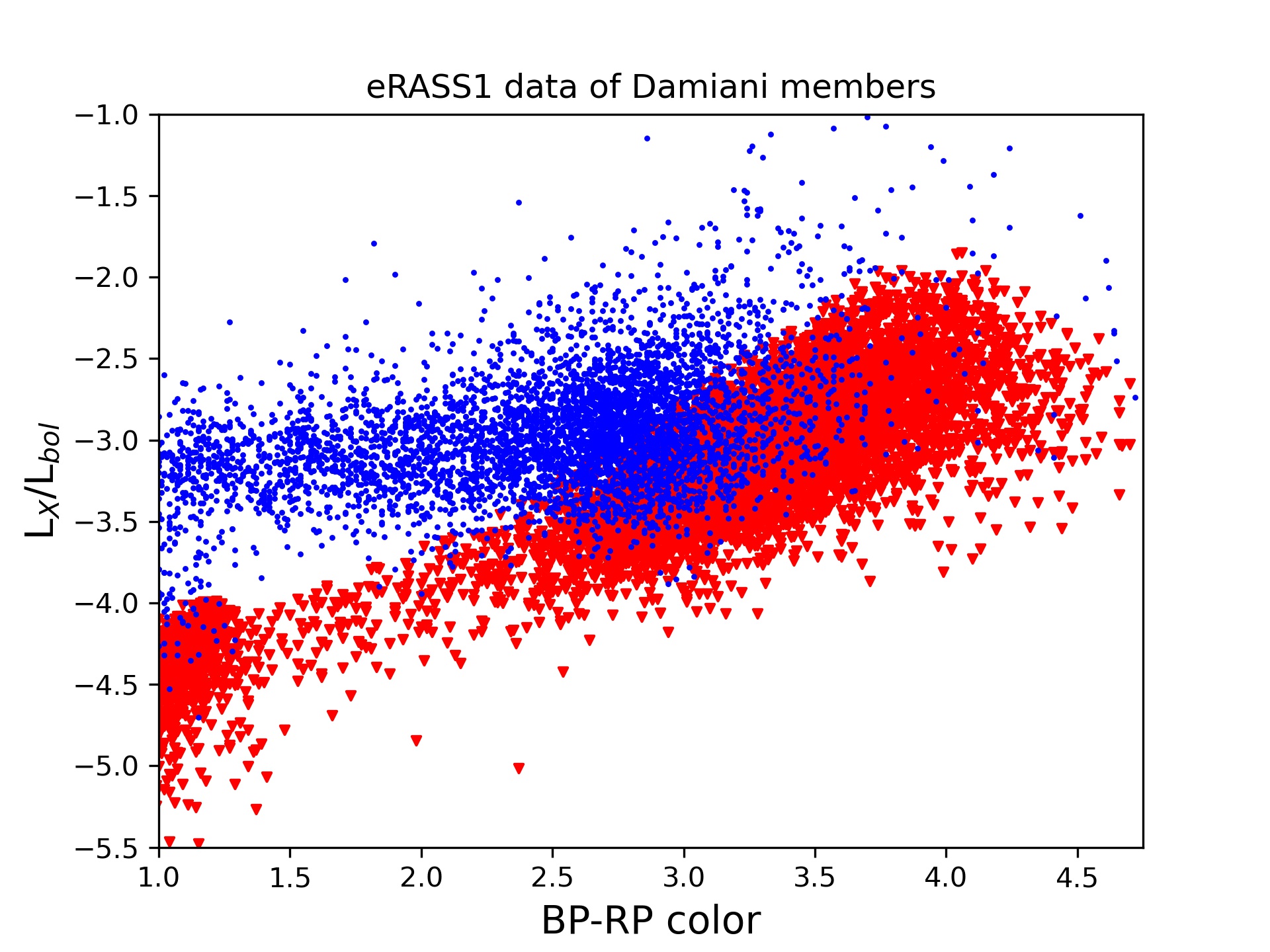}  
\caption{  L$_X$/L$_{bol}$ ratios for Sco-Cen members selected by  \cite{2019damiani}. Blue data points refer to 
eRASS1 detected Sco-Cen member candidates, and red data points to X-ray upper limits.} 
\label{scocen_damiani}  
\end{figure}

\subsubsection{Comparison to the membership lists derived by \cite{2019damiani} }
\label{sec_damiani}

We next compared the eROSITA detected X-ray sources in the Sco-Cen region to the membership list derived by \cite{2019damiani} on the basis
of {\it Gaia} DR2 data alone, focusing here again on the low-mass population of Sco-Cen. For the selection procedures applied by
\cite{2019damiani}, we refer to Sec.~\ref{sec_rev}. Specifically, \cite{2019damiani} selected only objects as members that
were located inside the red polygon displayed in Fig.~\ref{scocen_pecaut}.  A cross-check reveals that a substantial fraction of the 
stars proposed by  \cite{2019damiani} as members is in fact detected by eROSITA, but another large fraction is not.
In Fig.~\ref{scocen_damiani} we plot for the detected Damiani members the derived  L$_X$/L$_{bol}$ ratios  versus BP-RP color
(blue data points in Fig.~\ref{scocen_damiani}).  For the undetected Damiani members,  we can only estimate upper limits
to the X-ray fluxes, which we obtained by considering the minimally detected X-ray fluxes (from more than 700000 eRASS1 sources)
as a function of ecliptic latitude to account for the changing sensitivity of the eRASS survey.
Given the large size of Sco-Cen, this minimum flux varies substantially because the whole Sco-Cen complex extends from low (US) to rather high ecliptic latitudes (LCC).
The upper limits computed in this fashion  are plotted as red data points in Fig.~\ref{scocen_damiani}.  
The figure shows that
the L$_X$/L$_{bol}$ ratios of the detected members are almost exclusively at the saturation level of $\sim$ 10$^{-3}$, while
the upper limit values lie well below the detection values for stars with BP-RP colors $<$ 2.5. 
For redder objects, detections and nondetections
start to mix, and for objects with BP-RP $>$ 3.5, only very few detections have been obtained in eRASS1.  In the color range 
1 $<$ BP-RP $<$ 2.5, some gap appears between detections and upper limits, which probably means
that these nondetected stars should not be considered true Sco-Cen members.  For the red stars with BP-RP $>$ 3,  we can make
no real statement about the membership status of the undetected stars, but we note that with the future surveys
eRASS2-eRASS8, the sensitivity will probably to increase by some factors. This leads us to expect a more or less complete
detection of Sco-Cen members down to BP-RP $\sim$ 3.5, provided that these objects exhibit  saturated X-ray emission.

\begin{figure}  [bht]
\centering  
\includegraphics[scale=0.5]{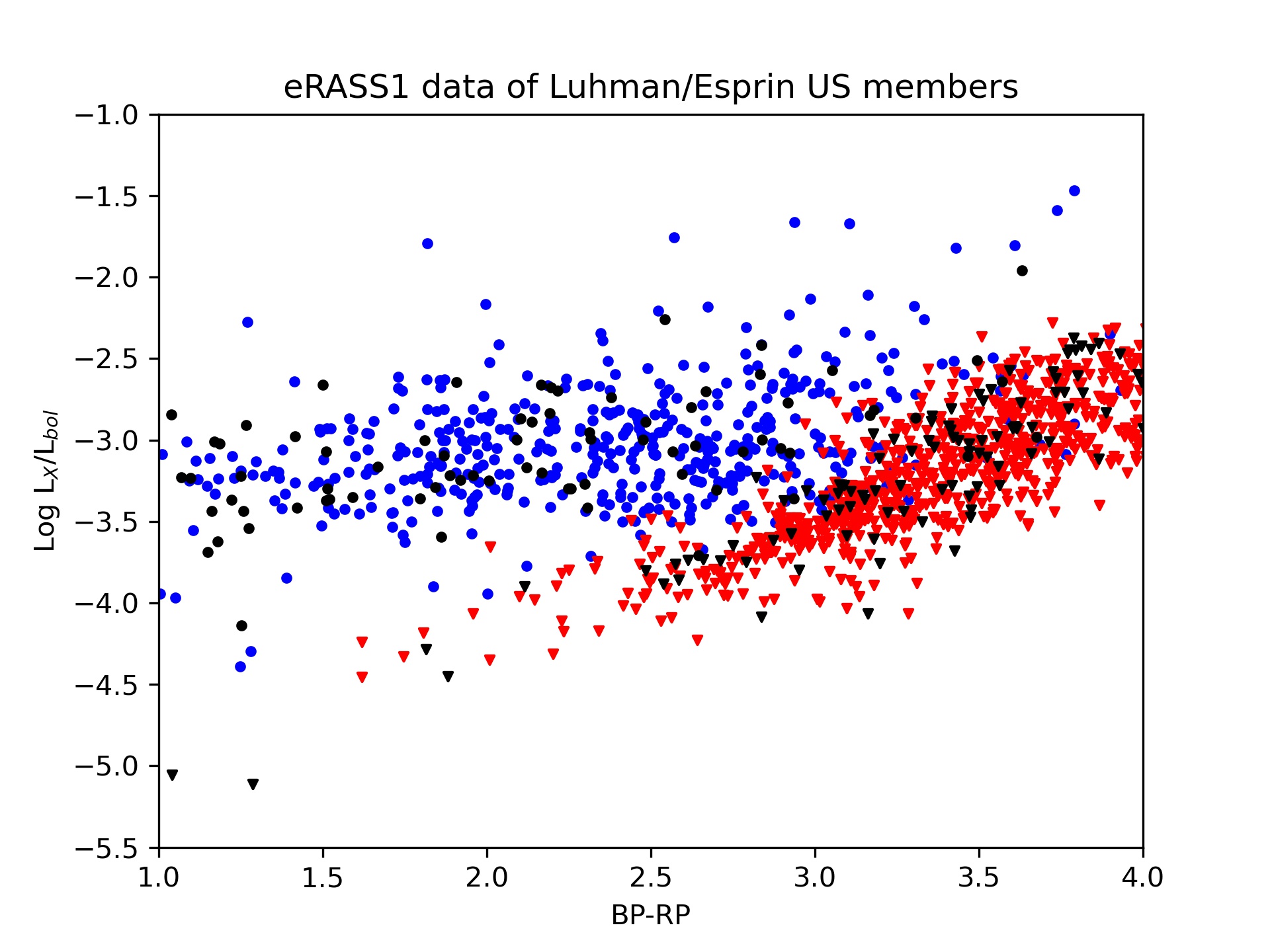}  
\caption{  L$_X$/L$_{bol}$ ratios for Sco-Cen members selected by  \cite{2020luhman}. Blue data points refer to 
eRASS1 detected Sco-Cen accepted member candidates, and black points to detected but rejected members.  Nondetections
are plotted as upper limits for accepted members (red data points) and rejected members (black data points).} 
\label{scocen_luhman}  
\end{figure}

\begin{figure*}  [t]
\centering  
\includegraphics[scale=1.2]{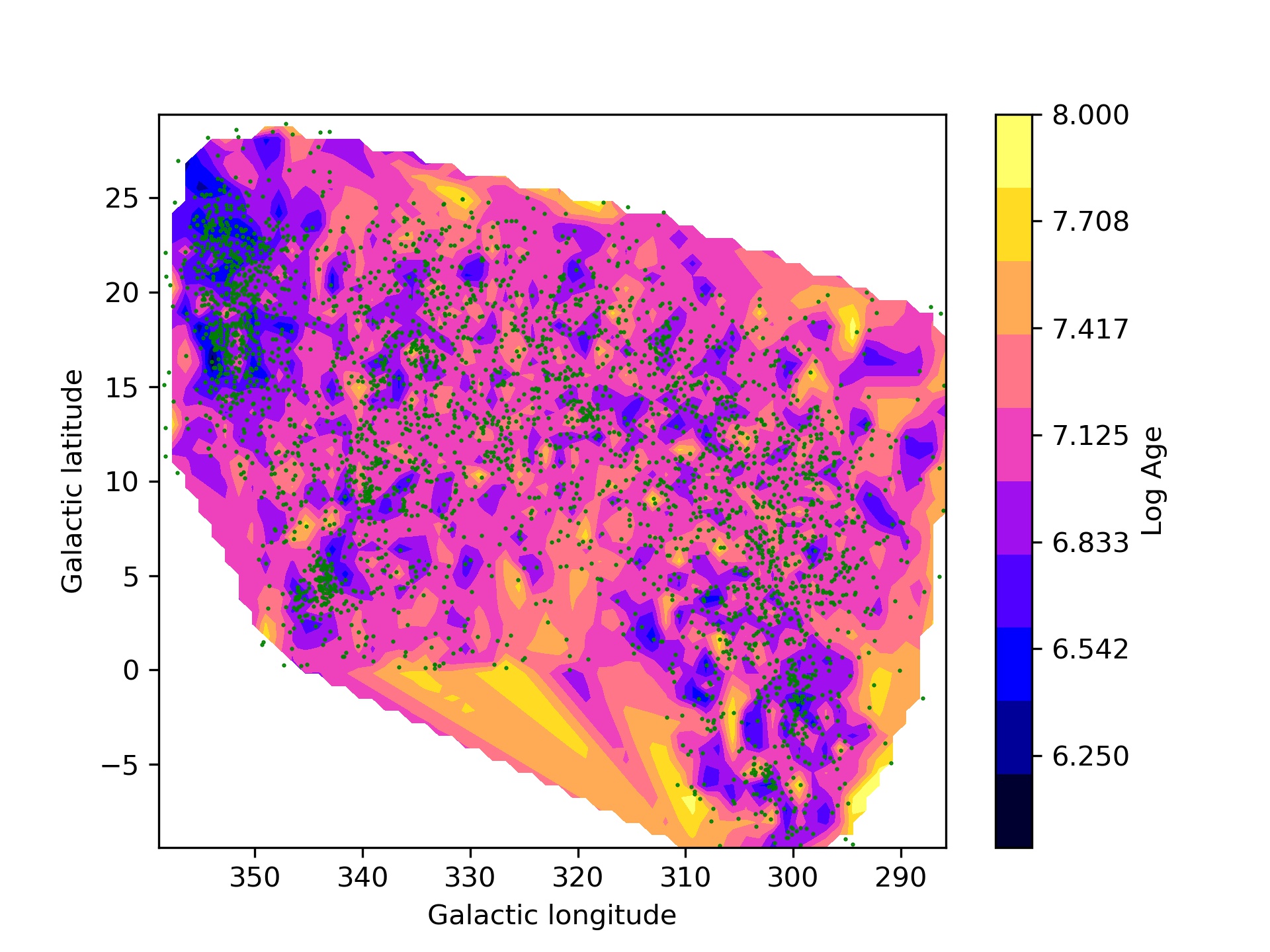}  
\caption{Mean (logarithmic) age map of the whole Sco-Cen association. Blue represents younger ages,
red older ages, and green dots denote the positions of late-type low-mass stars.}  
\label{sco_cenagemap}  
\end{figure*}

\subsubsection{Comparison to the membership lists derived by \cite{2020luhman} }

Finally, we used the US membership list derived by  \cite{2020luhman}  to compare with to our eRASS1 data.   \cite{2020luhman} pursued
a similar approach to that followed by  \cite{2019damiani}, but that their mixture model describes the proper motion
distribution statistically and not with a somewhat arbitrarily specified polygon (cf. Fig.~\ref{scocen_pecaut}). 
Because we are primarily interested in late-type stars, we again extracted (from Tab.~4 in \cite{2020luhman}) the stars with BP-RP $>$ 1, but 
removed objects without {\it Gaia} measured parallaxes, were fainter than  G = 18.4 mag, and were clearly located below the
US main sequence.  In this fashion,  we obtained a sample of 1568 stars, 224 of which were rejected by \cite{2020luhman} as members. Luhmann \&\ Esplin (2020)
provided a detailed description of their selection and rejection procedure.  

In the eRASS1 data we find 507 detections in
the accepted Sco-Cen members (detection rate: 38\%), and 108 detections in the rejected Sco-Cen members (detection rate: 48\%).
It appears strange that the detection rate for nonmembers is higher than that measured for members.
The reason for this oddity becomes apparent from Fig.~\ref{scocen_luhman}, where we plot  eROSITA measured L$_X$/L$_{bol}$ ratios for 
low-mass Sco-Cen members of \cite{2020luhman}.   Fig.~\ref{scocen_luhman} shows that the vast majority of undetected but accepted members consists of
rather red objects with BP-RP $>$ 3, while for stars with  BP-RP $<$ 2.5, the detection rate is very high, and the derived upper limits
are below the saturation level. This calls the membership status of these objects into question.

\subsubsection{Assessment of kinematic membership assignment}

An assessment of these results must remain ambiguous.   Apparently, the extraction region chosen by \cite{2019damiani}, which according to these authors should have a field star contamination of $<$ 3\%, contains quite a few stars with substantially reduced X-ray emission.  
This means that either we have to assume that the X-ray emission of these
bona fide low-mass Sco-Cen members is not necessarily saturated, or we have to assume that the contamination in the Damiani box is higher than 
estimated. Verifying the rotation periods of these stars would obviously be extremely useful.
We verified the properties of the membership sample derived by \cite{2016pecaut} and found that X-ray emission is indeed very common in their 
selected members, although X-ray emission was {\bf not} used directly as a filtering criterion by \cite{2016pecaut}. Many of the rejected members also show X-ray emission at comparable levels, however, and appear to be more widely distributed in the proper motion plane.  
Interestingly, quite a few stars rejected by \cite{2016pecaut} are 
located inside the Damiani box, but remain undetected by eROSITA.  Finally, quite a few stars rejected by \cite{2020luhman} are X-ray detected at
levels comparable to the level of accepted members.   In summary, it therefore appears to be unclear which criteria are best to determine the true members, but it also appears
that eROSITA detected X-ray emission is a necessary but possibly not sufficient criterion.

\section{Discussion and conclusions}
\label{sec:conc}

We have presented the results of the first eROSITA all-sky survey (eRASS1) in the Sco-Cen association
by cross-matching the eROSITA source catalog with {\it Gaia} EDR3.   Because 
both {\it Gaia} and eROSITA are ongoing survey missions,
our results must by necessity remain preliminary. They clearly demonstrate, however, that the combination of
eROSITA X-ray data and {\it Gaia} astrometry, kinematics and photometry will eventually allow a
comprehensive identification  of the low-mass stellar content of the huge Sco-Cen association. This
achievement will certainly form  a landmark for all studies of star-forming regions.
Even now we have a membership census comprising several thousand objects, and
with the eROSITA and {\it Gaia} surveys completed, this number will again increase by some factors.

With these data, it is clearly of interest to consider the change in the age distribution of stars throughout the huge Sco-Cen association.
Following \cite{2016pecaut}, we can construct an age map of the Sco-Cen association by computing the
mean (logarithmic) age of all bona fide Sco-Cen members as a function of celestial position presented in
Fig.~\ref{sco_cenagemap}. Our age map was derived exclusively from low-mass stars and looks qualitatively similar to the age map presented 
by  \cite{2016pecaut} ( in their Fig.~9).  The mean age decreases with galactic longitude, with the youngest
ages found in the US region. This result is well known.  Our eROSITA maps shows blue spots, however, that is, young stars 
younger than 10~Myr are sprinkled almost throughout the whole Sco-Cen region.  Our results therefore appear to
call into question a sequential star formation in the Sco-Cen region, and no unique age can be attributed
to the individual Sco-cen subgroups.

Assigning membership to stellar clusters and associations has been and still is a notoriously difficult task, and 
X-ray emission has so far played only a minor role in this process.
While traditionally, kinematic information (kinematics led to the discovery of
the Sco-Cen association in the first place) was widely used for membership assignment, 
the new ongoing {\it Gaia} and eROSITA missions provide high-precision CMDs and X-ray flux measurements
throughout the whole Sco-Cen region.   Our admittedly preliminary results demonstrate that X-ray emission must 
figure very importantly in the membership assigning process because
eROSITA is sensitive enough to detect essentially the whole low-mass stellar
population in the nearby Sco-Cen association when we make reasonable assumptions
about the nature of stellar X-ray emission.  We therefore argue that detected X-ray emission 
definitely must  be a necessary criterion for membership assignment. 

If we also use X-ray emission as a sufficient membership criterion,   we encounter the conundrum that some of these bona fide 
X-ray selected members, which are cospatial with the proper Sco-Cen members, violate the usually applied kinematic 
membership criteria and
are therefore not formal kinematic members.  The percentage of this diffuse population (in velocity space)
increases from US to UCL and LCC.  The stars contained in this population are X-ray sources, and
judging from their CMD positions, are also young, at least in their great majority (see Fig.~\ref{sco_cenhrd} ), but older on average than the compact population. Furthermore, their tangential positions
(see Fig.~\ref{scocentangvel} ) are similar to those of the kinematic members, but they are displaced
by about 10 km/s predominantly along the right ascension direction.  It thus unlikely that
this diffuse population has no physical connection to the compact population, but it is unclear
whether we see, for example, ejected members or the relics of an older generation of stars, or if
the origin of this population is unrelated to the Sco-Cen association.

 It is doubtful whether the assignment of a unique age to
the various Sco-Cen subgroups is a meaningful concept.  While there are definite
peaks in the age distribution of our bona fide Sco-Cen members, the age distributions
are quite broad and may also be interpreted as the result of various episodes
of star formation that are spread out in time.

A very substantial amount of optical follow-up work is clearly called for:
For thousands of very likely Sco-Cen members,
photometry and spectroscopy is required to determine the rotation periods of the objects,
check for the presence of disks around the stars, and search for signatures of binarity.  
Future releases of {\it Gaia} data will provide information on binarity, and
the classical tools of photometry and spectroscopy can be used to examine the Sco-Cen
stars for the presence of companions.  In essentially
all of these objects do we expect to find lithium absorption, and it would be extremely interesting to
determine the true amount of the spread in the lithium equivalent widths of the these stars.

Furthermore, these stars all have chromospheres that need to be characterized. Their relation to the detected X-ray 
emission needs to be studied.  Because all these stars show 
X-ray emission at the saturation level, we anticipate
strong chromospheric emission that should be easily detectable, for example,
in the cores of the Ca~II~H\&K lines where only very little photospheric emission
is expected given the low photospheric temperatures of these stars.  Clearly, the young stellar
population in the Sco-Cen region is likely to become a gold mine for ongoing or future
large spectroscopic surveys with multi-object spectrographs such as
GALAH \citep{2015desilva} or 4MOST \citep{2019dejong}.

Finally, the Sco-Cen stars
provide a golden sample to search for young planetary systems.   To elucidate the difference
between the compact and diffuse populations, spectroscopic surveys may reveal
differences between these populations. The accuracy of age assignment by CMD
especially for low-mass stars also needs to be investigated.   Once the complete set of
eROSITA survey data are available,
the expectation is that the luminosity and initial
mass functions of the various Sco-Cen subgroups can be determined down to M6 and
possibly further.
In summary, the new
data that are gathered by {\it Gaia} and eROSITA and the ensuing ground-based follow-up will
therefore certainly lead to a significantly improved  understanding of the star formation history in the
Sco-Cen association.

\begin{acknowledgements}
 
This work is based on data from eROSITA, the primary instrument aboard SRG, a joint
Russian-German science mission supported by the Russian Space Agency
(Roskosmos), in the interests of the Russian Academy of Sciences 
represented by its Space Research Institute (IKI), 
and the Deutsches Zentrum f\"ur Luft- und Raumfahrt (DLR). 
The SRG spacecraft was built by Lavochkin Association (NPOL) 
and its subcontractors, and is operated by NPOL with support from
IKI and the Max Planck Institute for Extraterrestrial Physics (MPE). 
The development and construction of the eROSITA X-ray instrument was
led by MPE, with contributions from the Dr.\ Karl Remeis Observatory 
Bamberg \& ECAP (FAU Erlangen-N\"urnberg), the University of Hamburg 
Observatory, the Leibniz Institute for Astrophysics Potsdam (AIP), 
and the Institute for Astronomy and Astrophysics of the University 
of T\"ubingen, with the support of DLR and the Max Planck Society. 
The Argelander Institute for Astronomy of the University of Bonn and 
the Ludwig Maximilians Universität Munich also participated in the science 
preparation for eROSITA.
The eROSITA data used for this paper were
processed using the eSASS/NRTA software system developed by the
German eROSITA consortium. 
This work has made use of data from the European Space Agency (ESA)
mission {\it Gaia} (\url{https://www.cosmos.esa.int/gaia}), processed by
the {\it Gaia} Data Processing and Analysis Consortium (DPAC,
\url{https://www.cosmos.esa.int/web/gaia/dpac/consortium}). Funding
for the DPAC has been provided by national institutions, in particular
the institutions participating in the {\it Gaia} Multilateral Agreement.
This research has also made use of the SIMBAD database,
operated at CDS, Strasbourg, France.
We acknowledge discussions with
Drs. C.~Bell and H.~Zinnecker, which very much helped to improve
this paper. 

\end{acknowledgements}

\bibliographystyle{aa}
\bibliography{bibfile}

\newpage
\onecolumn
\appendix
\section{eROSITA data for low-mass stars in the Sco-Cen association}
\label{apptable}

In Tab.~\ref{samplestars} we present the new eROSITA data for X-ray detected low-mass stars in the 
Sco-Cen region. All objects shown in Fig.~\ref{sco_cenhrd} are listed.  We specifically provide
the eROSITA names (in Col. 1) and the Gaia EDR3 entries (in Col. 2) with which the X-ray sources are associated.
Columns 3 and 4 provide the eROSITA X-ray coordinates ($\alpha$ and $\delta$ in degrees), and Col. 5 contains the distance
(in arcsec) between X-ray and Gaia position. Column 6 gives the eRASS1 count rate, Col. 7 the
probability for the correctness of the association between X-ray and optical source, and the flag in Col. 8
indicates the subregion in which the respective source is located.  We point out that these results are
derived from SASS processing 946, which is not the final version of the X-ray data.  The printed table only lists
some exemplary entries sorted by decreasing measured count rate. The table in its entirety 
is available electronically at Centre de Donn\'ees astronomiques de Strasbourg (CDS).

While a detailed comparison of the 6190 X-ray sources listed in Tab.~\ref{samplestars},  
for example, with the SIMBAD data base, is far beyond
the scope of this paper, it is of interest to investiagte the very brightest sources as a sanity check of our identification
scheme: 
The brightest source (1eRASS J113929.2-652348) is the RS~CVn type variable 12~Mus (g~=~4.9), 
and the second brightest source (1eRASS J160901.8-390512) is the source V908~Sco (g~=~13.6, spectral type M5.5),
a classical T Tauri star in the Lupus SFR \citep{2013galli}, which remained undetected in the ROSAT all-sky survey. Its high eRASS1 
X-ray flux is quite puzzling.
The third brightest source (1eRASS J134401.1-612157) is the well-known RS~CVn system V851~Cen (g~=~7.4). 
The fourth brightest source (1eRASS J121733.6-675739) is $\epsilon$ Mus  (g~=~2.7). This detection is entirely due to optical contamination, 
a topic addressed in Sec.~\ref{optcont}. The
eighth brightest source source (1eRASS J160538.0-203947) is again an active M star (g~=~11.9, spectral type M0) that was not detected
in the ROSAT all-sky survey.  This shows that while it might naively be assumed that the brightest stellar eROSITA sources are all
contained in the ROSAT all-sky survey, this is not the case due to the large variability of the X-ray sky.  As a
consequence, it is also clear that a study similar to the one presented here but with any of the subsequent eRASS surveys is
likely to yield a substantial number of late-type low-mass Sco-Cen members that are not contained in Tab.~\ref{samplestars}.

\begin{table*} [h]
\footnotesize
\begin{tabular}[h!]{llccccll}
\hline
\noalign{\smallskip}

SRG Name & Gaia ID & $\alpha$  & $\delta$ & Match & Rate    & Prob. & Group \cr
                   &              & (deg)       & (deg)     & (arcsec) & (cts/s) &          & \cr
\noalign{\smallskip}
\hline
\noalign{\smallskip}

eRASSU J113929.2-652348 & Gaia EDR3 5236523877939131008 & 174.8717 & -65.3969 & 3.7 & 4.92 10$^{1}$ & 0.86 & UCL \cr
eRASSU  J160901.8-390512 & Gaia EDR3 5997410491550194816 & 242.2577 & -39.0867 & 0.6 & 1.29 10$^{1}$ & 1.00 & LCC \cr
eRASSU  J134401.1-612157 & Gaia EDR3 5865871998528696576 & 206.0048 & -61.3658 & 2.3 & 1.02 10$^{1}$ & 0.99 & UCL \cr
eRASSU  J121733.6-675739 & Gaia EDR3 5859405805013401984 & 184.3902 & -67.9610 & 0.5 & 9.02 10$^{0}$ & 1.00 & UCL \cr
eRASSU  J114451.8-643853 & Gaia EDR3 5332613291464347392 & 176.2162 & -64.6482 & 1.8 & 7.59 10$^{0}$ & 1.00 & UCL \cr
eRASSU  J153701.9-313640 & Gaia EDR3 6208381582919629568 & 234.2582 & -31.6112 & 1.8 & 7.05 10$^{0}$ & 1.00 & LCC \cr
eRASSU  J152024.1-303733 & Gaia EDR3 6208166357818567680 & 230.1008 & -30.6260 & 1.7 & 6.48 10$^{0}$ & 1.00 & LCC \cr
eRASSU  J160538.0-203947 & Gaia EDR3 6244019542218710912 & 241.4087 & -20.6632 & 0.9 & 6.20 10$^{0}$ & 1.00 & US \cr
eRASSU  J125608.9-612724 & Gaia EDR3 6055724499479516544 & 194.0372 & -61.4568 & 2.9 & 6.15 10$^{0}$ & 1.00 & UCL \cr
eRASSU  J121131.5-581652 & Gaia EDR3 6071307396762179200 & 182.8814 & -58.2814 & 1.5 & 5.41 10$^{0}$ & 1.00 & UCL \cr
eRASSU  J164944.6-362422 & Gaia EDR3 6019764628108588160 & 252.4358 & -36.4063 & 0.3 & 5.37 10$^{0}$ & 1.00 & LCC \cr
eRASSU  J160335.4-224557 & Gaia EDR3 6243154501445899264 & 240.8977 & -22.7658 & 0.7 & 5.35 10$^{0}$ & 1.00 & US \cr
eRASSU  J121858.0-573719 & Gaia EDR3 6071727753787733120 & 184.7420 & -57.6221 & 1.0 & 4.73 10$^{0}$ & 1.00 & UCL \cr
eRASSU  J121906.8-630951 & Gaia EDR3 6054173981974483328 & 184.7784 & -63.1644 & 2.4 & 4.25 10$^{0}$ & 0.99 & UCL \cr
eRASSU  J154131.2-252036 & Gaia EDR3 6238094097954087808 & 235.3801 & -25.3435 & 0.3 & 4.11 10$^{0}$ & 1.00 & US \cr
eRASSU  J153129.6-302154 & Gaia EDR3 6209492506280547200 & 232.8734 & -30.3651 & 0.4 & 4.07 10$^{0}$ & 1.00 & LCC \cr
eRASSU  J125928.2-555113 & Gaia EDR3 6061481095694462976 & 194.8676 & -55.8536 & 2.1 & 3.81 10$^{0}$ & 1.00 & UCL \cr
eRASSU  J122720.1-581835 & Gaia EDR3 6059526198381792000 & 186.8339 & -58.3099 & 2.8 & 3.74 10$^{0}$ & 0.97 & UCL \cr
eRASSU  J145944.5-342547 & Gaia EDR3 6203845650778424960 & 224.9357 & -34.4298 & 1.1 & 3.69 10$^{0}$ & 1.00 & LCC \cr
eRASSU  J143548.1-524057 & Gaia EDR3 5897665511103933824 & 218.9507 & -52.6826 & 1.9 & 3.67 10$^{0}$ & 1.00 & LCC \cr
eRASSU  J124807.5-443915 & Gaia EDR3 6138425847095084288 & 192.0313 & -44.6543 & 2.8 & 3.36 10$^{0}$ & 0.98 & UCL \cr
eRASSU J162740.1-242205 & Gaia EDR3 6049153921054413440 & 246.9173 & -24.3682 & 2.0 & 3.27 10$^{0}$ & 1.00 & US \cr
eRASSU  J161028.8-221348 & Gaia EDR3 6242860961897271168 & 242.6202 & -22.2301 & 0.3 & 3.14 10$^{0}$ & 1.00 & US \cr
eRASSU  J165451.2-245534 & Gaia EDR3 4113215554323332096 & 253.7135 & -24.9263 & 1.5 & 3.13 10$^{0}$ & 1.00 & US \cr
eRASSU  J122420.5-544353 & Gaia EDR3 6076080033122502400 & 186.0857 & -54.7316 & 0.4 & 3.11 10$^{0}$ & 1.00 & UCL \cr

\noalign{\smallskip}
\hline

\end{tabular}
\vskip 0.25cm
\caption{\label{samplestars} Basic X-ray properties of the sample stars; columns 1 and 2 list the eROSITA and {\it Gaia} EDR3 names, columns 3 and 4 give 
the X-ray positions, column 5 the offset between X-ray and optical position, column 6 the measured eRASS1 count rate, columns 7 the association probability
and column 8 is a flag indicating the subgroup the object belongs to. }
\normalsize
\end{table*}

\section{Optical contamination}
\label{optcont}

The eROSITA detectors are prone to optical contamination.  Optical photons also lead to the
creation of charges in the CCD pixels of the detectors, and for an optically brighter source, the charge deposited in this way during the accumulation time of 50~ms can be as large as that deposited
by an actual X-ray photon.   In this way, optical photons can produce false X-ray photons.
For a pointing observatory such as {\it XMM-Newton,} it is therefore
possible to choose blocking filters of various thicknesses depending on the expected optical
brightness of the X-ray sources.  For a scanning mission such as eROSITA, this is not feasible, and the optical contamination is unavoidable.  To illustrate the effects of optical contamination,
we plot in Fig.~\ref{optcont} the recorded X-ray
count rate as a function of apparent g magnitude for all stars listed in Tab.~\ref{samplestars} .  
For the brighter stars one notices a ``cutoff'' in the
sense that at a given optical apparent magnitude a minimal count rate exists, indicated by the
dotted black line in Fig.~ref{optcont}; at the bright end lies the source  $\epsilon$ Mus,
discussed in Sec.~\ref{apptable}.  
All sources located in the vicinity of the black line
in the diagram of the count rate versus apparent magnitude need to be treated with extreme caution because
they are very likely due to optical contamination and not to true X-ray flux.  Fig.~\ref{optcont}
also shows that for the vast bulk of the sources reported in this paper, optical contamination is not
an issue, however.

\begin{figure}  [bht]
\centering  
\includegraphics[scale=0.5]{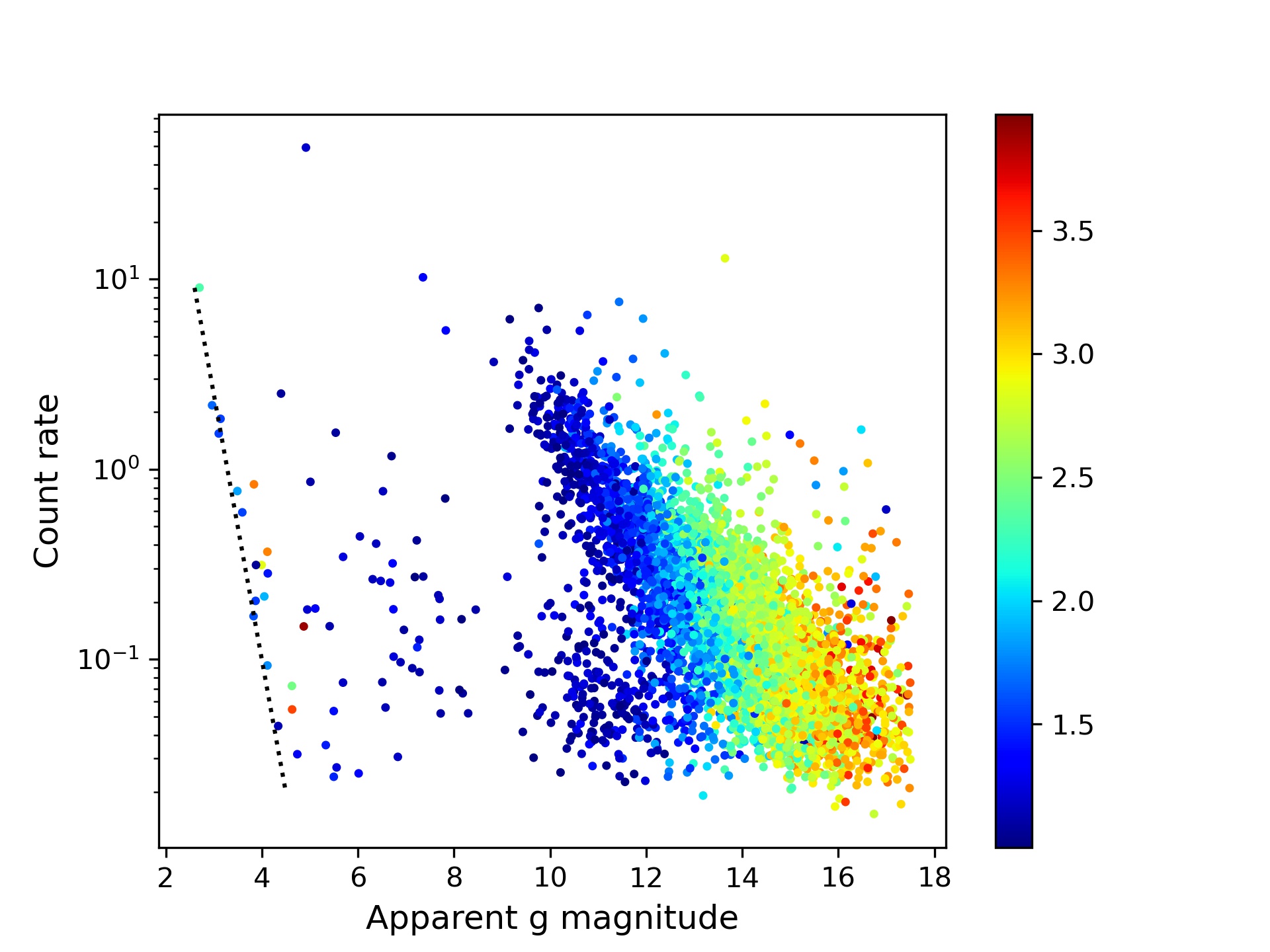}  
\caption{eROSITA count rate vs. apparent g magnitude for all stars listed in Tab.~\ref{samplestars}.  The color-coding
reflects the BP-RP color of the respective objects, and the dotted black line indicates the level of optical contamination.  Stars
located close to this line may be substantially optically contaminated. See text for more details.} 
\label{optcont}  
\end{figure}
\end{document}